\documentclass{amsart}
\usepackage{graphicx}
\usepackage{color}
\usepackage{tikz}
\usepackage{graphicx}
\usepackage{graphicx}
\usepackage{amssymb}
\usepackage{amsfonts}
\usepackage{amsmath}
\usepackage{amsthm}
\newtheorem{theo}{Theorem}[section]
\newtheorem{Lemma}[theo]{Lemma}
\newtheorem{Definition}[theo]{Definition}

\newtheorem{Corollary}[theo]{Corollary}
\newtheorem{Remark}{Remark}[part]

\newcommand{\be}{\begin{equation}}
\newcommand{\ee}{\end{equation}}

\numberwithin{equation}{section}

\begin{document}

\title{Exponential Utility with Non-Negative Consumption}
 \author{\small{Rom Aviv (Roman Muraviev)} \\ \\
Actuarial Research Center, Department of Statistics,  University of Haifa, Israel\\ \\
 Argsup Capital Ltd, Tel-Aviv, Israel}%
\address{Actuarial Research Center, Department of Statistics, University of Haifa, Mount Carmel, 31905 Haifa, Israel.
 \newline
Argsup Capital Ltd., 1 Cremieux St, Tel-Aviv 6425901, Israel.
\newline
 {E-mail: rom@argsupcapital.com}}

 \date{June 24, 2019}
\begin{abstract}
We offer mathematical tractability and new insights for a framework of exponential utility with non-negative consumption,
a constraint often omitted in the literature giving rise to economically unviable solutions. Specifically, using the Kuhn-Tucker theorem and the notion of aggregate state price density (Malamud and Trubowitz (2007)), we provide a solution to this problem in the setting of both complete and incomplete markets (with random endowments). Then, we exploit this result to provide an explicit characterization of complete market heterogeneous equilibria. Furthermore, we construct concrete examples of models admitting multiple (including infinitely many) equilibria. By using Cramer's large deviation theorem, we study the asymptotics of equilibrium zero coupon bonds. Lastly, we conduct a study of the precautionary savings motive in incomplete markets.

\end{abstract}

\maketitle
\markboth{R. Aviv (Muraviev)}{Exponential Utility with Non-Negative Consumption}
\renewcommand{\theequation}{\arabic{section}.\arabic{equation}}
\pagenumbering{arabic}
%\end{frontmatter}

\section{Introduction}
Utility maximization constitutes a primary field of research in financial mathematics,
since it offers a well-posed methodology for studying decision making under uncertainty.
%Utility maximization in the setting of decision making under uncertainty
%is a central topic both in mathematical finance and financial economics.
The most prevalently used classes of utilities to depict preferences are
CRRA (constant relative risk-aversion) and CARA (constant absolute risk-aversion).
Focusing on the latter class, we investigate various aspects of the corresponding maximization problem
with a rather economically viable constraint: \textit{consumption levels are not allowed
to take negative values.}

%We investigate various aspects of the discrete-time exponential utility maximization problem from
%consumption, where a significant and quite novel dimension introduced here (when comparing to other papers dealing with
%CARA preferences and consumption) is the constraint that feasible
%consumption policies \textit{are not permitted to be negative}.

CARA (or, exponential) utilities have attracted a great deal of attention during the last several decades. Concretely,
they have been employed in diverse applications in finance and insurance such as indifference pricing (Frei and Schweizer (2008),
Henderson (2009), and Frei et al. (2011)), incomplete markets with portfolio constraints  (Svensson and Werner (1993))
and asset pricing in equilibrium (e.g. Christensen et al. (2011a, b)).
Many researchers allow for negative consumption when studying exponential utility maximization (see e.g. Caballero (1990), Christensen et al. (2011a, b), and Svensson and Werner (1993)).
%An essential drawback challenging the classical paradigm of exponential utility is the fact that negative consumption
%
%Economically, and intuitively, this is a vague assumption
%which cannot be reasonably justified, but is rather made to allow a simplified framework when solving analytically
%the corresponding problem. Namely, consumption of goods over a life-cycle cannot be apparently measured in negative units,
%since it models unrealistic situations when for instance one is capable of consuming a negative amount of bread.
Solely for the purpose of alleviating complications caused by randomness of market models,
negative levels of consumption lack any other adequate justification. More bluntly, the negativity of consumption is
artificially introduced to obtain tractable solutions in the associated consumption/investment maximization problem.
Most authors, are aware of this drawback. For instance, Caballero (1990) writes:
"This paper specializes to this type of preferences (exponential) in
spite of some of its \textbf{unpleasant} features like the possibility of \textbf{negative
consumption}. (Unfortunately, explicitly imposing non-negativity constraints
impedes finding a tractable solution.)". Nonetheless, it is demonstrated in the current paper that it makes sense to impose this non-negativity
constraint. Not only that it genuinely refines the model, but it also allows us to solve
in closed-form a variety of economic problems.

We outline now the content and contributions of this work, and present its links to the existing literature.
First, by using some ideas from convex analysis, we solve the constrained (i.e. with non-negative consumption) exponential utility maximization problem
in a complete market setting. We show that the solution is equal to the non-negative part of the solution
associated with the unconstrained problem (see also Cox and Huang (1989) for a related problem in a continuous-time framework).
Next, we use the Kuhn-Tucker theorem to solve the preceding maximization problem in a setting of incomplete markets on a finite probability space,
and express the solution in terms of the aggregate state price density introduced by Malamud and Trubowitz (2007).
%In particular, we show that the transparent relation between the constrained and unconstrained solution in complete markets
%setting is violated for incomplete markets.
Then, we turn to analyzing heterogeneous equilibria in the framework of complete markets.
There is a vast body of literature studying equilibrium asset prices with heterogeneous investors
(see e.g. Mas-Colell (1986), Karatzas et al. (1990, 1991), Dana (1993a, 1993b), Constantinides and Duffie (1996),
Malamud (2008)). We express the equilibrium state price density as a 'non-smooth' sum over indicators depending on the agents' characteristics and the
total endowment of the economy. In effect, our finding is one of very few examples of a closed-form characterizations
the equilibrium state price density. Next, we concentrate on non-uniqueness of equilibria.
Non-uniqueness is usually anticipated in this type of models, due to
the fact that the Inada condition is not fulfilled,
leading to a violation of the so-called gross-substitution property (see Dana (1993a)), which would guarantee uniqueness.
However, we are not aware of any papers (apart from Malamud and Trubowitz (2006)) that construct other examples
of multiple equilibrium state price densities in a risk-exchange economy.
We then shift our attention to studying long-run limits (see e.g. Wang (1996), Lengwiler (2005), and Malamud (2008))
of zero coupon bonds, whose price is determined endogenously in \textit{heterogeneous equilibrium.}
The main tool we employ for the preceding problem is Cramer's large deviation theorem.
Finally, we explore the precautionary savings motive (see Kimball (1990) for a comprehensive introduction)
in incomplete markets with exponential preferences.
We verify that un-insurable future income forces an investor to save more (or equivalently, consume less) in the present.
We examine this phenomenon in a rather general stochastic framework of incomplete market (market of type $\mathcal{C}$, see
Malamud and Trubowitz (2007)), whereas most classical papers on this topic (see e.g. Dreze and Modigliani (1972)
and Miller (1976)) consider markets consisting of riskless bonds.

The paper is organized as follows. In Section 2 we introduce the model.
In Section 3 we establish a solution to the exponential utility maximization problem
with non-negativity constraint in both complete and incomplete markets.
Section 4 analyzes the associated heterogeneous equilibrium for complete markets.
In Section 5 we study the long-run behavior of equilibrium zero coupon bonds.
Finally, in Section 6 we explore the precautionary savings motive in incomplete markets.

\section{Preliminaries}
We consider a discrete-time market with a maturity date $T$.
In Sections 2-4 and Section 6, we choose $T \in \mathbb{N}$. In Section
5, we set $T = \infty.$ The uncertainty in our model is captured by a
probability space $ \left( \Omega,
\mathcal{F} , P \right) $ and a filtration $ \mathcal{F}_0 =
\left\{ \phi , \Omega \right\} \subseteq \mathcal{F}_1 \subseteq
... \subseteq \mathcal{F}_T = \mathcal{F}.$ Adaptedness and predictability
of stochastic processes is always meant with respect to the filtration $ ( \mathcal {F}_{k} )_{k=0,...,T},$
unless otherwise stated. We use the notation $\mathbb{R}_{+}=[0,\infty)$ and $\mathbb{R}_{++}=(0,\infty).$
There is no-arbitrage in the market which consists
of $n$ risky stocks with prices processes $(S^{j}_k)_{k=0,...,T}$, $j=1,...,n$ and
one riskless bond paying an interest-rate $r_k$, at each period $k=1,...,T.$
Each price process $(S^{j}_k)_{k=0,...,T}$ is non-negative and adapted,
and the interest rate process $(r_k)_{k=1,...,T}$ is non-negative and predictable.
The economy is inhabited by $N$ (types of) individuals, labeled by $i=1,...,N.$
Each agent $i$ receives a random income $ ( \epsilon^{i}_{k}
)_{k=0, ... , T}$, which is assumed to be non-negative and adapted.
The preferences of each agent $ i $ are characterized by
an impatience rate $\rho_i \geq 0$ and an exponential utility function $u_{i}(x) = - \exp (  - \gamma_i x) $,
\textit{defined on} $ \mathbb{R}_{+}$, for a given level of risk aversion
$ \gamma_i > 0$.
The underlying utility maximization problem from consumption of each agent $i$ is formulated as
\begin{equation}
\sup_{ (c_0 , ... , c_T) \in \mathcal{B}_i} \sum_{ k = 0 }^{ T } - e^{ - \rho_i k }  E \left[  \exp  \left( -
\gamma_i c_k \right)  \right],
\label{utility maximization}
\end{equation}
where $(c_k)_{k=0,...,T}$ is a consumption stream lying in a certain set of budget constraints $\mathcal{B}_i.$
As described below, we tackle problem (\ref{utility maximization}) in both complete and incomplete markets,
under different assumptions on the model and the budget set $\mathcal{B}_i.$
However, we preserve the same notations as above for both settings without mentioning it explicitly.
The market models introduced below are standard and mainly adapted from Duffie (2001).

\textit{2.1. Complete markets.} When dealing with complete markets,
the probability space is allowed to be infinite\footnote{Therefore, the amount of securities
completing the market might be infinite as well.}.
The budget set $\mathcal{B}_i$ in this setting consists of all
non-negative adapted processes $c_k$ that satisfy the equation
\begin{equation}
\sum_{k=0}^{T} E \left[ \xi_k c_{ k } \right] = \sum_{k=0}^{T} E \left[ \xi_k \epsilon^{i}_{ k } \right].
\label{constraint}
\end{equation}
Here, $(\xi_k)_{k=0,...,T}$ is the unique positive and normalized ($\xi_0=1$) state price density (SPD) of the market,
i.e.,
\[
S^{j}_k \xi_k =  E \left[ S^{j}_{k+1} \xi_{k+1}  | \mathcal{F}_k \right]  \ , \ j=1,...,n,
\]
and
\[
\xi_k =  E \left[ \xi_{k+1} ( 1 + r_{k+1} )   | \mathcal{F}_k \right],
\]
for $k=0,...,T-1$. Let us stress that the completeness of the market implies that stock prices and portfolios
are not directly involved in the corresponding utility maximization problem (\ref{utility maximization}). In particular,
the budget constraints are specified through the unique SPD (see (\ref{constraint})).
We introduce now the standard notion of equilibrium in the framework of a risk-exchange economy (see e.g. Malamud (2008)),
in which portfolios and specific stocks are omitted, due to the \textit{completeness} of the market
\begin{Definition}
An \textit{equilibrium} is a pair of processes $ ( c^{i}_{k} )_{ k= 0 , ... , T; i
= 1 , ... , N } $ and $ ( \xi_{k} )_{ k= 1 , ... , T } $ such that:
\newline
(a) The process $ ( \xi_{k} )_{ k= 1 , ... , T } $ is a SPD and $( c ^{i}_{k} )_{ k= 0 , ... , T } $ is the
optimal consumption stream of each agent $i$ (i.e., solving (\ref{utility maximization})).
\newline
(b) The market clearing condition
 \begin{equation}
\sum_{i=1}^{N} c^{i}_{k} = \epsilon_k :=  \sum_{ i=1}^{N}
\epsilon^{i}_{k}, \label{market_clearing}
\end{equation}
%\indent ${}$
holds for all $ k=0 , ... , T$.
\label{equilibrium_def}
\end{Definition}
\textit{2.2. Incomplete markets.} We assume that the probability space is finite. Incompleteness of markets is modeled standardly
by allowing infinitely many SPDs.
For each $k=0,...,T$, we denote by $L^{2}(\mathcal{F}_k)$ the Hilbert space of all $\mathcal{F}_k-$measurable
random variables, endowed with the inner product $\langle X , Y \rangle = E [ X Y] $, $X , Y \in L^{2}(\mathcal{F}_k).$
Each agent $i$ selects a portfolio strategy $(\pi^{j}_k)_{k=0,...,T-1},$ $j=1,...,n$, and $(\phi_k)_{k=0,...,T-1}$. Here,
$\pi^{j}_k$ and $\phi_k$ are $\mathcal{F}_k-$measurable and denote the shares invested in asset $j$ and the riskless bond
at period $k$, respectively. We set $\pi^{j}_{-1} = 0,$ $j=1,...,n$, $\phi_{-1}=0$,
$\pi^{j}_T = 0,$ $j=1,...,n$ and $\phi_T=0$. The last two assumptions formalize the convention that
no trading is executed in the last period $T$. For each $k=1,...,T$, we denote by
$$
\mathcal{L}_k = \left\{ \sum_{j=1}^{n} \pi^{j}_{k-1} S^{j}_k  +
\pi_{k-1} (1 + r_k) \bigg| \phi_{k-1}, \pi^{j}_{k-1} \in L^{2} \left( \mathcal{F}_{k-1} \right) \ , \ j=1,...,n \right\}
$$
the \textit{wealth space} at time $k$,
and remark that $ L^{2} \left( \mathcal{F}_{k-1} \right) \subseteq \mathcal{L}_k \subseteq L^{2} \left( \mathcal{F}_k \right) $.
By Lemma 2.5 in Malamud and Trubowitz (2007) there exists a unique normalized SPD $(M_k)_{k=0,...,T}$ such that $M_k \in \mathcal{L}_k,$
for all $k=1, ... ,T,$ called \textit{the aggregate SPD}. For incomplete markets, the budget constraints are more sophisticated than
the single constraint (\ref{constraint}) arising in a complete market setting. Namely, the budget set $\mathcal{B}_i$
is composed of (see Section 2 in Malamud and Trubowitz (2007)) all non-negative adapted processes $(c_k)_{k=0,...,T}$ of the form
\begin{equation}
c_k =  \epsilon^{i}_k
+ \sum_{j=1}^{n} \pi^{j}_{k-1} S^{j}_k  +
\pi_{k-1} (1 + r_k)
- \sum_{j=1}^{n} \pi^{j}_{k} S^{j}_k
- \pi_{k},
\label{constr_1}
\end{equation}
for all $k=0,...,T,$ where $(\pi^{j}_k)_{k=0,...,T-1},$ $j=1,...,n$, and $(\phi_k)_{k=0,...,T-1}$ is a portfolio strategy.
Notice that (\ref{constr_1}) can be rewritten as
\begin{equation}
c_k =  \epsilon^{i}_k + W_k - E \left[ \frac{M_{k+1} }{ M_k } W_{k+1} \big| \mathcal{F}_k  \right] ,
\label{constr_2}
\end{equation}
where $W_k \in \mathcal{L}_k$, $k=1,...,T,$ and $W_{0}=W_{T+1}=0$.

\section{Optimal Consumption}

\subsection{Complete Markets}
We investigate the utility maximization problem (\ref{utility maximization}) in the framework
of complete markets, as specified in Subsection 2.1.
We make use of convex conjugates and other related
ideas from convex analysis to derive an explicit
formula for the optimal consumption stream in this setup.
As will be shown below, there is a link between the utility maximization problem
(\ref{utility maximization}) and the corresponding unconstrained
(allowing negative consumption) version of this problem. Hence, we first treat the latter case.
\begin{theo}
Assume that $ 0 < \sum_{k=0}^{T} E [ \xi_k \epsilon^{i}_k ] < \infty $ and
$ - \infty < - \sum_{k=0}^{T} E [\xi_k \log \xi_k ] ,$ for all $i = 1 , ... , N.$
Consider the utility maximization problem
\begin{equation}
\sup_{(c_0,...,c_T)} \sum_{k=0}^{T} - e^{- \rho_i k } E \left[ \exp \left( - \gamma_i c_k  \right)  \right],
\label{unconstrained_1}
\end{equation}
where $(c_k)_{k=0,...,T}$ is an adapted process (not necessarily non-negative)
that satisfies equation (\ref{constraint}). Then, there exists a unique solution given by
\begin{equation}
  \tilde{c}^{i}_{k} =  \tilde{c}^{i}_{k} ( \widetilde{\lambda} ) :=
 \frac{1}{
\gamma_i } \left( \log \left( \frac{ \gamma_i  }{ \widetilde{\lambda}  }
\right) - \rho_i k - \log \left( \xi_k  \right) \right) ,
\label{unconstrained_3}
\end{equation}
for all $k=0,...,T,$ where $\widetilde{\lambda}$ is a positive real number
specified uniquely by the equation
\begin{equation}
\frac{1}{
\gamma_i }
\sum_{k=0}^{T} E \left[ \xi_k
\left( \log \left( \frac{ \gamma_i  }{ \widetilde{\lambda}  }
\right) - \rho_i k - \log \left( \xi_k  \right) \right)
\right] = \sum_{ k=0 }^{T} E \left[ \xi_k \epsilon^{i}_k \right].
\label{unconstrained_2}
\end{equation}
\label{unconstrained}
\end{theo}

\textbf{Proof of Theorem \ref{unconstrained}.} First, observe that the function
$f_{i} : \mathbb{R}_{++} \to \mathbb{R}$, $f_i(\lambda) = \sum_{k=0}^{T} E \left[ \xi_k
\left( \log \left( \frac{ \gamma_i  }{ \lambda  }
\right) - \rho_i k - \log \left( \xi_k  \right) \right)
\right]$ is strictly monotone decreasing with $\lim_{\lambda \to 0^{+}}$ $ f_{i} ( \lambda )  = \infty$ and
$\lim_{\lambda \to \infty } f_i(\lambda) = - \infty$. Therefore, there exists a unique solution $\widetilde{\lambda}$
for equation (\ref{unconstrained_2}). Next, consider the Legendre transform $\widetilde{v}_i (y) : \mathbb{R}_{++} \to \mathbb{R} $
of the function $-e^{-\gamma_i x }$, given by $
\widetilde{v}_i (y) = \sup_{ x \in \mathbb{R} } \left( - e^{ - \gamma_i x } - x y  \right).
$
Denote $ \widetilde{I}_{i}(y):= \frac{1}{\gamma_i} \log \frac{\gamma_i}{y}, $ and note that
$\widetilde{v}_i (y) = - e^{ - \gamma_i \widetilde{I}_{i}(y) } - \widetilde{I}_{i}(y) y .$ Existence now follows from the inequality
\[
  - e^{ - \rho_i k}   e^{ - \gamma_i \widetilde{I}_{i} \left(  \widetilde{\lambda} e^{\rho_i k} \xi_k \right) }
  \geq
- e^{ - \rho_i k} e^{  - \gamma_i c_k }
+ \widetilde{\lambda} \xi_k \left( \widetilde{I}_i \left( \widetilde{\lambda} e^{\rho_i k} \xi_k  \right) - c_k  \right) ,
\]
which holds for all $\mathcal{F}_k-$measurable random variables $c_k$.
Uniqueness follows from the inequality $\widetilde{v}_i (y) > - e^{ - \gamma_i x } - x y $, satisfied for all
$ x \neq \widetilde{I}_{i}(y).$ $\qed$
${}$
\newline
\newline
Next, we solve the utility maximization problem with the non-negativity constraint.

\begin{theo}
Assume that $  \sum_{k=0}^{T} E \left[ \xi_k
\epsilon^{i}_k  \right]  > 0.$ Then, the constrained utility maximization
problem (\ref{utility maximization}) in the setting of a complete market admits a unique
solution given by
\begin{equation}
c^{i}_k = \left( \tilde{c}^{i}_{k} ( \lambda^{*} ) \right)^{+} = \frac{1}{
\gamma_i } \left( \log \left( \frac{ \gamma_i  }{ \lambda^{*}  }
\right) - \rho_i k - \log \left( \xi_k  \right) \right)^{+},
\label{optimal_consumption}
\end{equation}
where the constant $ \lambda^{*} $ is determined as the unique positive solution of the equation
\begin{equation}
\frac{1}{
\gamma_i }
\sum_{k=0}^{T} E \left[ \xi_k
\left( \log \left( \frac{ \gamma_i  }{ \lambda^{*}  }
\right) - \rho_i k - \log \left( \xi_k  \right) \right)^{+}
\right] = \sum_{ k=0 }^{T} E \left[ \xi_k \epsilon^{i}_k \right]
. \label{lambda_star}
\end{equation}
\label{general_probability_space}
\end{theo}
${}$
\newline
We prove first the following auxiliary lemma.

\begin{Lemma} Denote $I_{i}(y) = \frac{1}{\gamma_i} \left( \log \left( \frac{\gamma_i}{y} \right) \right)^{+} $, and
consider the function $ \psi^{i} : \mathbb{R}_{++} \to \mathbb{R}_{+} $ defined by
\[
\psi^{i} ( \lambda ) = \sum_{k=0}^{T} E \left[ \xi_k I_{i} ( \lambda e^{\rho_i k }  \xi_k  )  \right].
\]
Then, $ \psi^{i} ( \lambda) $ is a decreasing continuous function
of the following form: if $ \psi^{i} (b) > 0 $ for some $ b \in
\mathbb{R}_{++}$, then $ \psi^{i} ( a ) > \psi^{i} ( b ) $ for
all $ 0 < a <  b $. Furthermore, $ \lim_{ \lambda \to 0^{+} } \psi^{i}
( \lambda  ) = \infty $ and $ \lim_{ \lambda \to  \infty  }
\psi^{i} ( \lambda  ) = 0. $ \label{properties_function}
\end{Lemma}
${}$
\newline
\textbf{Proof of Lemma \ref{properties_function}.} First observe
that $ E \left[ \xi_k  I_{i} ( c  \xi_k  ) \right] <  \infty $
for all $c >0$, since
\[
E \left[ \xi_k  I_{i} ( c  \xi_k  ) \right] =  \frac{1}{ \gamma_i
} E \left[ \xi_k  \log \left( \frac{ \gamma_i }{ c \xi_k}
\right)   \mathbf{1}_{ \{ 1 \leq \frac{ \gamma_i  }{ c \xi_k }   \} }  \right] < 1 / c,
\]
which follows from $ \log t < t,$ holding for all $t \geq 1$. Therefore,
$\psi^{i} ( \lambda)$ is well defined for all $\lambda \in \mathbb{R}_{++}$.
Proving the continuity of $\psi^{i} $ is routine, and thus omitted.
%Let $ \lambda_n $ be a
%sequence such that $ \lambda_n \uparrow \lambda ,$ hence, $ \lim_{ n \to \infty} \xi_k I_{i} ( \lambda_n  $ $e^{\rho_i k }  \xi_k ) = \xi_k I_i ( \lambda
 %e^{\rho_i k }  \xi_k  ) $, $P-$a.s, and thus the dominated convergence theorem implies that
 %$ \lim_{ n \to \infty } \psi^{i}( \lambda_n) = \psi^{i}( \lambda).$
 %The same argument holds for a sequence $ \lambda_n \downarrow \lambda.$ Now, we treat the limits. Consider an arbitrary increasing sequence
 %$ \lambda_n \to \infty $, then, $ \xi_k I_i ( \lambda_n  e^{\rho_i k }   \xi_k  ) \to 0  $ and $ \xi_k I_i ( \lambda_1
 %e^{\rho_i k }   \xi_k  )  \geq \xi_k I_i ( \lambda_n  e^{\rho_i k }   \xi_k  )   $, for all $n$.
 %Therefore, the dominated convergence theorem implies that $ \lim_{ n \to \infty } \psi^{i}( \lambda_n ) = 0 $. Next, pick an arbitrary
 %sequence $ \lambda_n \downarrow 0$ and note that $ \lim_{ n \to \infty } \xi_k I_i ( \lambda_n  e^{\rho_i k }   \xi_k  )  =  \infty $, $P-$a.s.
% Therefore, Fatou's lemma implies that
%\[
%\liminf_{ n \to \infty } E \left[ \xi_k I ( \lambda_n  e^{\rho k }  \xi_k  )   \right] \geq  E \left[  \liminf_{ n \to \infty } \xi_k I
% ( \lambda_n  e^{\rho k }  \xi_k  )   \right] =  \infty.
%\]
%This shows that $ \lim_{ \lambda \to 0^{+} } \psi^{i} ( \lambda  ) =
%\infty. $ Next, note that $ \lambda \mapsto E \left[ \xi_k I_{i}
%\left( \lambda e^{ \rho_i k} \xi_k \right) \right] $ is a
%decreasing function for each $k=1,...,T$ since $ I_{i}$ is
%decreasing. Therefore, $\psi^{i} ( \lambda ) $ is decreasing.
Next, assume towards contradiction that $ \psi^{i}(b) > 0 $ and $
\psi^{i}(a) = \psi^{i}(b) $, for some $a < b.$ It follows that $ E \left[ \xi_k  I_{i} \left( a e^{
\rho_i k} \xi_k \right)   \right] = E \left[ \xi_k I_{i} \left( b
e^{ \rho_i k} \xi_k \right) \right],$ for each $k=0,...,T$. By
definition, $I_{i}$ is a strictly decreasing function on the interval $ (0,
\gamma_i ]$, thus
 $ I_{i}
\left( b e^{ \rho_i k} \xi_k \right)   <
I_{i} \left( a e^{ \rho_i k} \xi_k \right)$ holds
on the set $\{  b e^{ \rho_i k} \xi_k < \gamma_i \}$.
Since $\psi^i(b)>0$, it follows that there exists some $k \in \{ 0 , ... , T\} $
such that $P \left[ b e^{ \rho_i k} \xi_k < \gamma_i \right]  > 0.$
This is a contradiction, since $ \{ b e^{ \rho_i k} \xi_k < \gamma_i\} \subseteq
\{ a e^{ \rho_i k} \xi_k < \gamma_i \} $, and thus
\[
 E \left[ \xi_k  I_{i} \left( a e^{
\rho_i k} \xi_k \right) 1_{ \left\{
b e^{ \rho_i k} \xi_k < \gamma_i \right\} }  \right]
<
 E \left[ \xi_k I_{i} \left( b
e^{ \rho_i k} \xi_k \right)
1_{ \left\{ b e^{ \rho_i k} \xi_k < \gamma_i \right\} }
 \right],
\]
completing the proof. $ \qed$
\newline
\newline
\textbf{Proof of Theorem \ref{general_probability_space}}. %First,
%observe that Lemma \ref{properties_function} yields the existence
%of a unique solution to equation (\ref{lambda_star}) denoted by $
%\lambda^{*} > 0.$ Next, consider the Legendre transform of the function
%$- e^{- \gamma_i x }$ with the domain of definition $\mathbb{R}_{+}$:
%\[
%v_{i}(y) = \sup_{x \in \mathbb{R}_{+} } \left(  - e^{- \gamma_i x } - x y \right),
%\]
%for all $y > 0.$ Note that $v_{i}(0)=0$ and recall that
%$I_i(y) = \frac{1}{\gamma_i} \left( \log \frac{\gamma_i}{y} \right)^{+}$.
%Observe that $ \frac{ \partial }{ \partial x } ( u_{i} (x) - xy  )  = \gamma_i e^{ - \gamma_i x } - y.$ Therefore,
%if $ y < \gamma_i $, then $ v_i (y) = u_i \left( \frac{1}{ \gamma_i } \log \left( \frac{ \gamma_i  }{ y }   \right)
% \right) - \frac{y}{ \gamma_i } \log \left( \frac{ \gamma_i  }{ y }   \right). $ If $ y \geq \gamma_i $, then $v_i(y) = u_i ( 0).$
%Thus, we can rewrite $ v_i (y) = u_i ( I_i (y) ) - I_i (y)  y$.
The proof makes use of Lemma \ref{properties_function} and follows the lines of the proof of Theorem \ref{unconstrained}. $ \qed$
\newline
\newline
For sufficiently large endowments, and under some
boundness assumptions on the SPD, the optimal
consumption stream is strictly positive.

\begin{Corollary}
Assume that $   \xi_{k} < C $ and $ \epsilon^{i}_k > \frac{1}{
\gamma_i } \left(  \log \left( \frac{ C }{  \xi_k }  \right) +
\rho_i \left( T- k  \right)  \right),$ $P-$a.s., for all
$k=0,...,T,$ and some constant $C>0$. Then, the
optimal consumption stream of the $i-$th agent (see (\ref{optimal_consumption}))
is strictly positive and given by
\begin{equation}
c^{i}_k =  \frac{1}{ \gamma_i  } \left(
\frac{ \sum_{l=0}^{T} E \left[ \xi_l \left( \gamma_i \epsilon^{i}_{l} + \log \xi_l + \rho_i l \right)  \right]   }{
\sum_{l=0}^{T} E \left[ \xi_l \right] }  - \rho_i k  - \log  \xi_k    \right) > 0, \label{pos_consumption}
\end{equation}
for all $k=0,...,T$.
\label{corollary}
\end{Corollary}

\subsection{Incomplete Markets}
In the current section we deal with incomplete markets and random endowments (see Subsection 2.2).
In this setting, we provide an explicit construction of the optimal consumption for
the utility maximization problem (\ref{utility maximization}). The methods employed here
rely on the Kuhn-Tucker theorem and the notion of aggregate SPD (defined in Subsection 2.2) introduced by
Malamud and Trubowitz (2007).
\begin{theo}
For an investor solving the utility maximization problem (\ref{utility maximization}) in an incomplete market,
the optimal consumption stream $(\widehat{c}_k)_{k=0,...,T}$ is uniquely determined through the following scheme:
\begin{equation}
P^{k}_{\mathcal{L}} \left[  \frac{ e^{ - \rho_i k } \gamma_i \exp ( - \gamma_i \widehat{c}_k )  + \lambda_k }{
e^{ - \rho_i (k-1) } \gamma_i \exp ( - \gamma_i \widehat{c}_{k-1} )  + \lambda_{k-1}  } \right] = \frac{M_k}{M_{k-1}},
\label{incomplete3}
\end{equation}
for all $k=1,...,T,$ where $(\lambda_k)_{k=0,...,T}$ is a non-negative adapted process satisfying
\[
\lambda_k \widehat{c}_k = 0,
\]
for all $k=0,...,T.$ Here, $(M_k)_{k=0,...,T}$ and $P^{k}_{\mathcal{L}}$ stand for the aggregate SPD and the orthogonal projection
on the space $\mathcal{L}_k$, respectively (see Subsection 2.2).
\label{incomplete1}
\end{theo}
\textbf{Proof of Theorem \ref{incomplete1}.}
Consider the function
\[
 l^{i}( \pi, \phi, \lambda_0 ,  ..., \lambda_T ) =
\]
\[
\sum_{k=0}^{T} e^{- \rho_i k} E \left[  - \exp \left( - \gamma_i \left( \epsilon^{i}_k + \sum_{j=1}^{n} \pi^{j}_{k-1} S^{j}_k
+ \phi_{k-1} (1 + r_k )
- \sum_{j=1}^{n} \pi^{j}_k S^{j}_k   - \phi_k \right) \right) \right]
\]
\[
-
\sum_{k=0}^{T} E \left[ \lambda_k
\left( \epsilon^{i}_k + \sum_{j=1}^{n} \pi^{j}_{k-1} S^{j}_k
+ \phi_{k-1} (1 + r_k )
- \sum_{j=1}^{n} \pi^{j}_k S^{j}_k   - \phi_k \right)
\right],
\]
where $ \phi = (\phi_k)_{k=0,...,T-1}, \pi = (\pi^{j}_k)_{k=0,...T-1}$, $j=1,...,n,$
is a portfolio strategy and $\lambda_k \in L^{2} \left( \mathcal{F}_k \right)$, $k=0,...,T$.
Since our probability space is finite, we can employ the Kuhn-Tucker theorem. Namely, by differentiating the
function $l^{i}$ with respect to the partial derivatives $\phi_k, \pi^{j}_k$, $k=0,...,T$, $j=1,...,n$, and
equalizing the resulting expression to $0$, we get that the optimal consumption stream $(\widehat{c}_k)_{k=0,...,T}$
is determined by requiring that the process
\[
\left(  e^{ - \rho_i k } \gamma_i \exp \left( - \gamma_i
\widehat{c}_k \right) + \lambda_k \right)_{k=0,...,T}
\]
is a SPD and $\lambda_k \widehat{c}_k = 0,$ $k=0,...,T$, for some non-negative adapted process $(\lambda_k)_{k=0,...,T}$.
Finally, Lemma 2.5 in Malamud and Trubowitz (2007) yields the validity of
(\ref{incomplete3}). Uniqueness follows from strict concavity. $\qed$

\begin{Remark}
The Kuhn-Tucker theorem could not be applied directly for complete markets,
since we allowed for arbitrary probability spaces.
\end{Remark}

\textbf{Example: one-period incomplete markets.}
We fix $T=1$ and consider a market of type $\mathcal{C}$, introduced by Malamud and Trubowitz (2007).
That is, we assume that $\mathcal{L}_1 = L^{2} \left( \mathcal{H}_1 \right)$ and thus $P^{\mathcal{L}}_1
\left[  \cdot \right] = E \left[ \cdot \big| \mathcal{H}_1 \right] ,$ where $\mathcal{H}_1$ is a sigma-algebra satisfying
$ \mathcal{F}_0 \subseteq \mathcal{H}_1 \subseteq \mathcal{F}_1 .$ Let $\widehat{c}_0$ and  $\widehat{c}_1$ denote the optimal
consumption stream (we drop the index $i$). As above, $\lambda_0$ and $\lambda_1$ stand for the multipliers.
Recall that by (\ref{constr_2}), we have $\widehat{c}_1 = \epsilon^{i}_1  + \widehat{W}_1 $, where $\widehat{W}_1 \in \mathcal{L}_1$.
Next, by Theorem \ref{incomplete1} we get
\[
 \exp \left( - \gamma_{i} \widehat{W}_1 \right)
=
\frac{ M_1 \left(
\gamma_{i} \exp ( - \gamma_{i} \widehat{c}_{0} ) + \lambda_{0} \right)   -
E \left[ \lambda_1 \big| \mathcal{H}_1 \right] }{ e^{-\rho_{i} } \gamma_{i}
E \left[ \exp \left( - \gamma_{i} \epsilon^{i}_1 \right) \big| \mathcal{H}_1 \right] },
\]
hence
\begin{equation}
\widehat{c}_1 = \epsilon^{i}_1 + \widehat{W}_1 =
 \frac{1}{\gamma_{i}} \log \left( \frac{ e^{\gamma_{i} \epsilon^{i}_1 } E \left[ e^{- \gamma_{i} \epsilon^{i}_1 } \big| \mathcal{H}_1 \right] }{
\left( \exp \left( \rho_{i} - \gamma_i \widehat{c}_{0} \right) + \lambda_{0} \frac{ e^{\rho_{i} } }{\gamma_{i}} \right) M_1 -
\frac{e^{\rho_{i} } }{ \gamma_i }
E \left[ \lambda_1 \big| \mathcal{H}_1 \right]
 }  \right).
\label{consumption_1per}
\end{equation}
Now, denote $ \lambda = \exp \left( \rho_{i} - \gamma_{i} c_0 \right)  + \lambda_0 \frac{e^{\rho_{i}}}{\gamma_{i}},
$ or equivalently $ \widehat{c}_0 = \frac{1}{\gamma_i} \log \left( \frac{\gamma_i }{ \lambda \gamma_{i} e^{-\rho_{i}} - \lambda_0 }  \right)$.
Recall that $\lambda_0 \widehat{c}_0=0,$ $\lambda_0 \geq 0$ and $\widehat{c}_0 \geq 0$. Therefore, we get
\begin{equation}
\widehat{c}_0 = \frac{1 }{ \gamma_i} \left(
\log \left( \frac{1 }{ \lambda }   \right) + \rho_i
\right)^{+}.
\label{c0_incomplete}
\end{equation}
Now, we claim that
\begin{align}
\widehat{c}_1 =
\frac{1}{\gamma_i} \log \left( \frac{ e^{\gamma_i \epsilon^{i}_1}  E \left[ e^{ - \gamma_i \epsilon^{i}_1 } \big| \mathcal{H}_1 \right] }{ \lambda M_1 } \right)
1_{ \{ essinf [ e^{\gamma_i \epsilon^{i}_1} | \mathcal{H}_1 ] E
[ e^{ - \gamma_i \epsilon^{i}_1} | \mathcal{H}_1 ]
> \lambda M_1 \} }
\label{c1_incomplete}
\\ +
 \frac{1}{\gamma_i}
\log \left( \frac{ e^{ \gamma_{i} \epsilon^{i}_1 } }{ essinf [ e^{  \gamma_{i} \epsilon^{i}_1 }  | \mathcal{H}_1 ] } \right)
1_{ \{
essinf [ e^{\gamma_{i} \epsilon^{i}_1} | \mathcal{H}_1 ] E
[ e^{ - \gamma_{i} \epsilon^{i}_1} | \mathcal{H}_1 ]
 \leq \lambda M_1 \} },
\nonumber
\end{align}
where $essinf[ e^{  \gamma_{i} \epsilon^{i}_1 } | \mathcal{H}_1]$ is the \textit{essential infimum} of the random variable $e^{  \gamma_{i} \epsilon^{i}_1 }$
conditioned on the sigma-algebra $\mathcal{H}_1$. To this end, assume first that $ e^{\gamma_i \epsilon^{i}_1} E \left[ e^{ - \gamma_i \epsilon^{i}_1} \big| \mathcal{H}_1 \right]
 >  \lambda M_1$.
Then, identity (\ref{consumption_1per}) yields $\widehat{c}_1 > 0 $ and thus $\lambda_1 =0,$ since $\lambda_1 \widehat{c}_1=0.$
Thereby, we have
$$ \lambda_1 = \Lambda 1_{ \left\{  e^{\gamma_i \epsilon^{i}_1} E \left[ e^{ - \gamma_i \epsilon^{i}_1} \big| \mathcal{H}_1 \right]  \leq
\lambda M_1 \right\}}, $$ for some $\mathcal{F}_1-$measurable non-negative random variable $\Lambda$. Now, the condition $\lambda_1 \widehat{c}_1 = 0$
can be rewritten as
\begin{equation}
 \Lambda 1_{ \left\{  e^{\gamma_i \epsilon^{i}_1} E \left[ e^{ - \gamma_i \epsilon^{i}_1} \big| \mathcal{H}_1 \right]  \leq
\lambda M_1 \right\}}
\times
\label{inc_crit}
\end{equation}
\[
  \log \left( \frac{ e^{\gamma_i \epsilon^{i}_1} E \left[ e^{ - \gamma_i \epsilon^{i}_1} \big| \mathcal{H}_1 \right] }{
\lambda M_1 -
\frac{e^{\rho_{i} } }{ \gamma_{i} }
E \left[
 \Lambda 1_{ \left\{  e^{\gamma_i \epsilon^{i}_1} E \left[ e^{ - \gamma_i \epsilon^{i}_1} \big| \mathcal{H}_1 \right]  \leq
\lambda M_1 \right\}}
  \big| \mathcal{H}_1 \right]
 }  \right) = 0 .
\]
Hence, if
$$ essinf [ e^{\gamma_i \epsilon^{i}_1} | \mathcal{H}_1] E \left[ e^{ - \gamma_i \epsilon^{i}_1} \big| \mathcal{H}_1 \right] > \lambda M_1,
$$
then $ \widehat{c}_1 = \frac{1}{\gamma_i} \log \left( \frac{ e^{\gamma_i \epsilon^{i}_1} E \left[ e^{ - \gamma_i \epsilon^{i}_1} \big| \mathcal{H}_1 \right] }{ \lambda M_1 } \right) .$
On the other hand, we claim that if
$$ essinf [ e^{\gamma_i \epsilon^{i}_1} | \mathcal{H}_1] E \left[ e^{ - \gamma_i \epsilon^{i}_1} \big| \mathcal{H}_1 \right]
\leq \lambda M_1,
$$
then
$$
  \log \left( \frac{ essinf \left[ e^{\gamma_i \epsilon^{i}_1 } \big| \mathcal{H}_1 \right] E \left[ e^{ - \gamma_i \epsilon^{i}_1} \big| \mathcal{H}_1 \right] }{
\lambda M_1 -
\frac{e^{\rho_{i} } }{ \gamma }
E \left[
 \Lambda 1_{ \left\{  e^{\gamma_i \epsilon^{i}_1} E \left[ e^{ - \gamma_i \epsilon^{i}_1} \big| \mathcal{H}_1 \right]  \leq
\lambda M_1 \right\}}
  \big| \mathcal{H}_1 \right]
 }  \right) = 0 .
$$
Assume that it is not the case. It follows that
\[
  \log \left( \frac{ e^{\gamma_i \epsilon^{i}_1} E \left[ e^{ - \gamma_i \epsilon^{i}_1} \big| \mathcal{H}_1 \right] }{
\lambda M_1 -
\frac{e^{\rho_{i} } }{ \gamma_i }
E \left[
 \Lambda 1_{ \left\{  e^{\gamma_i \epsilon^{i}_1} E \left[ e^{ - \gamma_i \epsilon^{i}_1} \big| \mathcal{H}_1 \right]  \leq
\lambda M_1 \right\}}
  \big| \mathcal{H}_1 \right]
 }  \right) > 0,
\]
and thus by (\ref{inc_crit}) we get
$$ \Lambda 1_{ \left\{  e^{\gamma_i \epsilon^{i}_1} E \left[ e^{ - \gamma_i \epsilon^{i}_1} \big| \mathcal{H}_1 \right]  \leq
\lambda M_1 \right\}} = 0.$$
By substituting it back, we get
$$
essinf [ e^{\gamma_i \epsilon^{i}_1} | \mathcal{H}_1] E \left[ e^{ - \gamma_i \epsilon^{i}_1} \big| \mathcal{H}_1 \right]
> \lambda M_1,$$
and this is a contradiction, proving the identity (\ref{c1_incomplete}). Finally, let us remark that $\lambda$ is
derived from the equation $\widehat{c}_0 + E \left[ M_1 \widehat{c}_1 \right] = \epsilon^{i}_0 + E \left[ M_1 \epsilon^{i}_1 \right].$
This closes the example. $\qed$
%\begin{Remark} For complete markets, the optimal consumption stream in the
%constrained case (non-negative consumption) is given in the form of a call-option
%on the unconstrained (possibly negative) optimal consumption stream with a strike
%being equal to zero (see (\ref{optimal_consumption})).
%In contrast to this, for incomplete markets this relation is no longer valid. Indeed,
%if we allow for negative consumption, one can show that the associated unconstrained optimal
%consumption stream is given by
%\[
% c'_1 =
%\frac{1}{\gamma_i} \log \left( \frac{ e^{\gamma_i \epsilon^{i}_1}  E \left[ e^{ - \gamma_i \epsilon^{i}_1 } \big| \mathcal{H}_1 \right] }{ \lambda' M_1 } %\right),
%\]
%and
%\[
% c'_0 = \frac{1}{\gamma_i} \left( \rho_i + \log \frac{1}{\lambda'} \right),
%\]
%where the constant $\lambda'$ is determined by the equation $ c'_0 + E \left[ M_1 c'_1  \right]
%= \epsilon^{i}_0 + E \left[ M_1 \epsilon^{i}_1  \right].$ Thus by comparing it to
%(\ref{c0_incomplete}) and (\ref{c1_incomplete}), we conclude that
%the structure of the solution presented in the complete setting disappears here.
%\end{Remark}

\section{Equilibrium}
\subsection{Existence and characterization}
In the present subsection we provide a closed-form formula for complete-market equilibrium SPDs
(see Definition \ref{equilibrium_def}), and prove existence.
%We adapt the notion of a complete market equilibrium as introduced in Subsection 2.1
%(see Definition \ref{equilibrium_def}). In the current section we show that there exists an
%equilibrium under the assumption that all endowments are
%positive. Moreover, we describe the set of all feasible
%equilibrium SPDs and the associated optimal
%consumption policies.
For this purpose we introduce the following
quantities. For each vector $ ( \lambda_1 , ... , \lambda_N) \in
\mathbb{R}^{N}_{++}$, we define
\begin{equation}
\beta_{i }(k) = \beta^{( \lambda_i )}_{i }(k) = \frac{
\gamma_i }{ \lambda_i e^{ \rho_i k } } , \label{beta}
\end{equation}
for all $i=1,...,N$ and all $k=0,...,T$. For a fixed $k=0,...,T,$
let $ i_1(k) , ... , i_N(k) $ denote the order statistics of
$\beta_1(k),...,\beta_{N}(k)$, that is, $ \{ i_1(k) , ... , i_N(k)
\} = \{ 1 , ... , N \} $ and $ \beta_{i_1 (k) }(k) \leq ... \leq
\beta_{i_N (k) }(k)$. We set $ \beta_{i_0 (k)}(k) = 0 $, for all
$k=0,...,T.$ With the preceding notations, we denote
\begin{equation}
\eta_{j}(k) = \eta^{( \lambda_1 , ... , \lambda_N)}_{j}(k) \
= \sum_{l=j+1}^{N} \frac{ \log \left( \beta_{ i_{l}(k)
}(k) \right) - \log \left( \beta_{ i_{j}(k) }(k) \right)  }{
\gamma_{ i_{l} (k) } } \geq 0.
\label{eta}
\end{equation}
Note that $ \eta_{0}(k) = + \infty $ and $ \eta_{N}(k) = 0,$ for
all $k=0 , ... , T.$ Lastly, we introduce a candidate for the
equilibrium SPD
\begin{equation}
 \xi_{k} \left( \lambda_1 , ... , \lambda_N \right) =
 \label{spd}
\end{equation}
\[
\sum_{j=1}^{N}  \left( \prod_{ l=j }^{ N } \beta^{ \left(
\sum_{m=j}^{N} ( \gamma_{ i_{l}(k) } / \gamma_{ i_m (k) } )
\right)^{-1}     }_{ i_{l}(k) }    \exp \left( -
\frac{\epsilon_k}{ \sum_{l=j}^{N} 1 / \gamma_{ i_l (k) }  }
\right) \right)  \mathbf{1}_{ \{ \eta_j(k) \leq \epsilon_k <
\eta_{j-1}(k)   \} } ,
\]
for all $k=0,...,T.$ Recall that $I_{i}(y) = \left( \log \frac{ \gamma_i}{y} \right)^{+} $.

\begin{theo}
Assume that $ \epsilon^{i}_k > 0 $ for each period $k$ and each agent $i$,
$P-$a.s. Then, there exists an equilibrium. Furthermore, every equilibrium SPD is given by
$ \left( \xi_k\left( \lambda^{*}_1 , ... , \lambda^{*}_N \right) \right)_{k=0,...,T}$,
where $ \lambda^{*}_1 , ... , \lambda^{*}_N \in \mathbb{R}_{++}$ are constants
solving the following system of equations
\begin{equation}
\sum_{k=0}^{T} E \left[ \xi_k ( \lambda_1 , ... , \lambda_N )
I_{i} ( \lambda_i e^{ \rho_i k } \xi_k ( \lambda_1 , ... ,
\lambda_N )  )    \right] = \sum_{ k = 0 }^{ T } E \left[  \xi_k
\left( \lambda_1 , \lambda_2 , ... , \lambda_N \right)
\epsilon^{i}_k \right],
\label{numbers}
\end{equation}
for $i=1,...,N$.
\label{het_equil1}
\end{theo}

\textbf{Proof of Theorem \ref{het_equil1}.} Let $(
c^{i}_{k})_{k=0,...,T}$ denote the optimal consumption stream of
agent $i$. Recall that by (\ref{optimal_consumption}) we have
$ c^{i}_{k} = I_{i} \left(  \lambda^{*}_i e^{ \rho_i k }
\xi_k \right)$ for some $\lambda^{*}_i>0$. Plugging this into the
market clearing condition (\ref{market_clearing}), we obtain that
the following holds in equilibrium:
\begin{equation}
\sum_{i=1}^{N} I_{i} \left(  \lambda^{*}_i e^{ \rho_i k }  \xi_k
\right) = \epsilon_k, \label{equation2}
\end{equation}
for all $k=0,...,T$. Here, $\lambda^{*}_1,...,\lambda^{*}_N$ are constants
that will be derived from the budget constraints in the sequel.
Using the explicit form of $I_i$, this is equivalent to
\[
\sum_{i=1}^{N} \log \left[ \left( \frac{ \gamma_i  }{  \gamma_i
\mathbf{1}_{ \left\{ \lambda^{*}_i e^{ \rho_i k }  \xi_k > \gamma_i
\right\} } +  \lambda^{*}_i e^{ \rho_i k }  \xi_k  \mathbf{1}_{
\left\{ \lambda^{*}_i e^{ \rho_i k }  \xi_k \leq \gamma_i \right\} }
}     \right)^{ 1 / \gamma_i }  \right] = \epsilon_k ,
\]
a further transformation yields,
\begin{equation*}
 \prod_{ i=1 }^{ N }  \left(   \gamma_i   \mathbf{1}_{ \left\{  \xi_k > \beta_i(k)  \right\} } +  \lambda^{*}_i e^{ \rho_i k }  \xi_k  \mathbf{1}_{ \left\{ \xi_k \leq \beta_i(k)   \right\} }    \right)^{ 1 / \gamma_i }   =  \prod_{i=1}^{N} \gamma^{ 1 / \gamma_i }_i  \exp( - \epsilon_k ),
\end{equation*}
where $ \beta_i(k) = \beta^{ ( \lambda^{*}_i ) }_i(k) $ was defined in (\ref{beta}). This is equivalent to
\begin{equation}
\sum_{j=1}^{N} Y^{(k)}_{j}  \mathbf{1}_{  \left\{ \beta_{ i_{ j-1
}(k) }(k)  <  \xi_k  \leq \beta_{ i_j(k) }(k)  \right\}  }  + \prod_{i=1}^{N} \gamma^{ 1 / \gamma_i }_i
\mathbf{1}_{ \left\{ \xi_k > \beta_{i_N (k) }(k)   \right\}  } = \prod_{i=1}^{N} \gamma^{ 1 / \gamma_i }_i
  \exp \left(  -  \epsilon_k \right) ,
\label{modif}
\end{equation}
where
\[
Y^{(k)}_j = \prod_{l=1}^{j-1} \gamma^{ 1 / \gamma_{ i_{l}(k) }  }_{ i_{l}(k) } \prod_{l=j}^{N} ( \lambda^{*}_{ i_{l}(k) } )^{ 1 / \gamma_{ i_{l}(k) }  }
 \exp{ \left( k \sum_{ l = j }^{ N } \frac{ \rho_{i_l (k) } }{ \gamma_{i_l (k) } }  \right)}  {\xi_k}^{ \sum_{ l = j }^{ N } \frac{1}{ \gamma_{ i_l(k) } }    } .
\]
The strict-positivity assumption on the endowments implies that
$ \xi_k \leq \beta_{ i_N} (k)$
holds $P-$a.s. Next, for each $k=0,...,T$, we have
\[
Y_{i}^{(k)} \mathbf{1}_{ \left\{ \beta_{ i_{ j-1 } (k) }(k) < \xi_k
\leq \beta_{ i_j (k) }(k) \right \} } =   \prod_{i=1}^{N} \gamma^{ 1 / \gamma_i }_i  \exp \left( -
\epsilon_k \right) \mathbf{1}_{ \left\{ \beta_{ i_{ j-1 } (k) }(k) <
\xi_k \leq \beta_{ i_j (k) }(k) \right \} },
\]
which implies that the following holds on each set $ \left\{
\beta_{ i_{ j-1 } (k) }(k) < \xi_k \leq \beta_{ i_j (k) }(k)
\right \}:$
\[
\xi_k = \prod_{ l=j   }^{  N }  (\beta_{i_l(k) }(k))^{ \left(  \sum_{ m = j }^{ N } \gamma_{ i_l(k) } / \gamma_{ i_m(k) }     \right)^{-1}   }    \exp \left(  -  \frac{ \epsilon_k }{\sum_{ l=j   }^{ N } 1 / \gamma_{ i_l(k) } } \right).
\]
In particular, one checks that $ \xi_k \leq \beta_{ i_j (k) }(k) $
is equivalent to $ \epsilon_k \geq \eta_j (k) $ and $ \beta_{ i_{
j-1 } (k) }(k) < \xi_k $ is equivalent to $ \epsilon_k <
\eta_{j-1} (k), $ where, $ \eta_j(k) = \eta^{\lambda^{*}_1 , ... ,
\lambda^{*}_N }_j (k) $ is given in (\ref{eta}). Now, one rewrites the above identity in terms of $
\lambda^{*}_1,...,\lambda^{*}_N$ and concludes that every
equilibrium SPD is of the form (\ref{spd}) for
some $\lambda^{*}_1,..., \lambda^{*}_N$.
Existence is standard and follows from Theorem
17.C.1 in Mas-Colell et al. (1995).
$\qed$
\newline
\newline
\textbf{Example: homogeneous economy.}  In an economy populated
only by agents of type $i$ that hold strictly positive endowment
streams $( \epsilon^{i}_k )_{k=0,...,T}$, there exists a unique (normalized)
equilibrium and the corresponding homogeneous SPD
process $ \{ \xi^{i}_k \}_{k=0,...,T}$ is given by
$$ \xi^{i}_k  = e^{ \gamma_i \left( \epsilon^{i}_0 - \epsilon^{i}_k   \right)  - \rho_i k}, $$ for all $k=1,...,T.$
The optimal consumptions obviously coincide with the endowments: $
c^{i}_k = \epsilon^{i}_k,$ for all $k=0,...,T$. $qed$

\subsection{Non-uniqueness of equilibrium}

\subsubsection{Non-uniqueness with positive endowments.}
The system of equations (\ref{numbers}) can admit multiple solutions, causing
non-uniqueness of equilibrium. This can be anticipated due to the
fact that the \textit{gross substitution} property (which is closely related to
the absence of the Inada condition for our exponential preferences; see Definition 3.1 in Dana (1993b))
is generally violated. Otherwise this would be sufficient for uniqueness.
 %Furthermore, the lack
%of the latter regularity features turns out to be crucial for
%uniqueness issues.
The next example demonstrates the existence of
multiple equilibria.
\newline
\newline
\textbf{Example: non-uniqueness of equilibrium.} We assume a one
period market with $\mathcal{ F}_0 = \mathcal{F}_1 = \{  \Omega,
\emptyset \}$. Consider two agents $i=1,2$ represented by $
u_{1}(x) =  u_{2}(x) = -e^{-x} $ and $\rho_1 = \rho_2 = 0$. The agents hold different
endowments $\epsilon^{1}_{0} , \epsilon^{1}_{1} $ and
$\epsilon^{2}_{0} , \epsilon^{2}_{1} ,$ respectively. Let $
\epsilon_0$ and $\epsilon_1$ denote the aggregate endowments. By
Theorem \ref{het_equil1}, every equilibrium state price density is
of the form
\[
\xi_1(x,y) = \frac{1}{ \min \{ x , y \} } \mathbf{1}_{ \{ \epsilon_1 < \log \frac{ \max \{ x , y  \} }{  \min \{ x , y \} } \} }  e^{ - \epsilon_1 } +
\frac{1}{ \sqrt{xy} } \mathbf{1}_{ \{ \log \frac{ \max \{ x , y  \} }{  \min \{ x , y \} } \leq \epsilon_1 \} }  e^{ -  \epsilon_1 /  2},
\]
where $x$ and $y$ are to be determined by the budget constraints. One checks that
\begin{equation*}
\xi_1  ( x , y  ) =
\begin{cases} \frac{1}{ e^{ \epsilon_1}  }  \frac{1}{x}  &\text{if $ 0 < x  < \frac{ y  }{ e^{ \epsilon_1 }  }  $,}
\\
\frac{1}{ \sqrt{ y } e^{ \epsilon_1 / 2 }  }  \frac{1}{ \sqrt x }
&\text{if $  y  e^{ - \epsilon_1 } \leq x \leq y  e^{ \epsilon_1 }
$,}
\\
\frac{1}{ y e^{ \epsilon_1} }  &\text{if $  x  >  y  e^{ \epsilon_1 }  $.}
\end{cases}
\end{equation*}
The positive arguments $ x , y $ solve equations (\ref{numbers})
which take the form:
\[
(1) \ \log ( 1 / y ) \mathbf{1}_{ \{  y \leq 1  \} } + \xi_1 ( x , y  ) \log \left( \frac{1}{  y \xi_1(x , y )  } \right) \mathbf{1}_{ \{ x \xi_1 (x , y ) \leq 1   \} }    = \epsilon^{1}_0 + \epsilon^{1}_1 \xi_1 ( x , y  ) ,
\]
\[ (2) \
\log ( 1 / x ) \mathbf{1}_{ \{  x \leq 1  \} } + \xi_1 ( x , y ) \log \left( \frac{1}{  x \xi_1 (x , y )  } \right) \mathbf{1}_{ \{ x \xi_1 (x , y ) \leq 1   \} }
= \epsilon^{2}_0 + \epsilon^{2}_1 \xi_1 ( x , y  ).
\]
Let us note that we work with a normalized state price density,
i.e., $ \xi_0 = 1.$ We denote
$$ h( x , y ) = \xi_1(x , y ) \left( \log \left( \frac{1}{  y \xi_1 (x , y )   } \right) \mathbf{1}_{ \{ x \xi_1 (x , y ) \leq 1   \} } - \epsilon^{1}_1 \right), $$ and
$$ g(x , y ) = \xi_1(x , y ) \left( \log \left( \frac{1}{  x \xi_1 ( x , y )  }  \right) \mathbf{1}_{ \{ x \xi_1 (x , y ) \leq 1   \} }   - \epsilon^{2}_1 \right) .$$ Observe that
\begin{equation*}
h(x , y) =
\begin{cases} - \epsilon_1^{1} \frac{1}{e^{ \epsilon_1 } }  \frac{1}{x}  &\text{if $ 0 < x  < \frac{ y  }{ e^{ \epsilon_1 }  }  $,}
\\
\frac{ 1  }{ \sqrt{ y  } e^{  \epsilon_1  /  2 }  }  \frac{1}{
\sqrt{x} } \left( \log \left(   \frac{ \sqrt{x} e^{ \epsilon_1  / 2 }  }{ \sqrt{ y } }  \right)   - \epsilon^{1}_1  \right)
&\text{if $  y  e^{ - \epsilon_1 } \leq x \leq y  e^{ \epsilon_1 }
$,}
\\
\frac{1}{ y e^{ \epsilon_1} } \left( \epsilon_1 - \epsilon^{1}_1   \right)  &\text{if $  x  >  y  e^{ \epsilon_1 }  $,}
\end{cases}
\end{equation*}
and
\begin{equation*}
g( x , y ) =
\begin{cases} - \epsilon_1^{2} \frac{1}{e^{ \epsilon_1 } }  \frac{1}{x}  &\text{if $ 0 < x  < \frac{ y  }{ e^{ \epsilon_1 }  }  $,}
\\
\frac{ 1  }{ \sqrt{ y  } e^{ - 1 / 2 \epsilon_1 }  }  \frac{1}{
\sqrt{x} } \left( \log \left(   \frac{ \sqrt{ y } e^{ 1 / 2
\epsilon_1 }  }{ \sqrt{ x } }  \right)   - \epsilon^{2}_1  \right)
&\text{if $  y  e^{ - \epsilon_1 } \leq x \leq y  e^{
\epsilon_1 }    $,}
\\
\frac{1}{ y e^{ \epsilon_1} } \left( \epsilon_1 - \epsilon^{2}_1   \right)  &\text{if $  x  >  y  e^{ \epsilon_1 }  $.}
\end{cases}
\end{equation*}
Hence, equations (1) and (2) from above can be rewritten as
\[
(1') \ \log ( 1 / y ) \mathbf{1}_{ \{ y \leq 1  \} }  +  h(x , y )   = \epsilon^{1}_0,
\]
\[
(2') \ \log ( 1 / x ) \mathbf{1}_{ \{ x \leq 1  \} }  +  g(x ,  y )   = \epsilon^{2}_0.
\]
We are going to construct two solutions $(x_1, y_1)$
and $( x_2 , y_2)$ such that $ y_l < 1 $ and $ 0 < x_l < y_l e^{ -
\epsilon_1 } ,$ for $l = 1 , 2.$ Set $ \epsilon^{1}_1 =
\epsilon^{2}_1 $, $ e^{ 2\epsilon^{1}_1 -1 } > \epsilon^{1}_1 $
and $ \epsilon^{1}_1 < e^{-1}$. We start by treating equation
$(2')$. Consider the function $ h(x) = \log( 1 / x ) - \frac{
\epsilon^{1}_1  }{ e^{ \epsilon_1 } } \frac{1}{x} $ on the
interval $[ 0 , y e^{ - \epsilon_1 } ] $, for arbitrary $y>0$.
Note that $ h'(x) > 0$ for $ x <
\frac{ \epsilon^{1}_1 }{ e^{ \epsilon_1 } }, $ and $ h'(x) < 0
$ for $ x
> \frac{ \epsilon^{1}_1 }{ e^{ \epsilon_1 } } $, which implies
that $ x_{max} = \frac{ \epsilon^{1}_1 }{ e^{ \epsilon_1 } }$ is a maximum of $h$.
Furthermore, $y_l$ (to be determined explicitly in the sequel) will satisfy
$ \epsilon^{1}_1  < y_l $, guaranteeing that the maximum is
indeed in the domain of definition of $h$, namely
 $ \frac{ \epsilon^{1}_1 }{ e^{ \epsilon_1 } } \in [0, y_l e^{
- \epsilon_1 }]$. Next, note that $ h ( \frac{ \epsilon^{1}_1
}{ e^{ \epsilon_1 } }  ) = \log \left( \frac{ e^{ \epsilon_1 }  }{
\epsilon^{1}_1}  \right) - 1 > 0, $ due to the assumption $ e^{
2\epsilon^{1}_1 -1 } > \epsilon^{1}_1 $. Now let $ \delta > 0$ be
some small quantity to be determined below. One can pick $
\epsilon^{2}_0$ such that the equation $ h(x) =\epsilon^{2}_0 $
has exactly two solutions $x_1$ and $x_2$ in the interval $ [
\frac{ \epsilon^{1}_1 }{ e^{ \epsilon_1 } }  - \delta , \frac{
\epsilon^{1}_1 }{ e^{ \epsilon_1 } } + \delta ].$ Now, equation
$(1')$ has two solutions denoted by $y_1$ and $y_2$ (depending on
$\epsilon^{1}_0$) corresponding to $x_1$ and $x_2$ that are given
by
$
y_{l} = \exp \left( - \epsilon^{1}_0 - \frac{ \epsilon^{1}_1  }{ e^{ \epsilon_1 }  } \frac{1}{x_l}  \right),
$
for $l=1,2$. Obviously, $ y_{1}, y_{2} < 1$. It is left to check that $ \max\{ x_l  , \frac{ \epsilon^{2}_1 }{ e^{ \epsilon_1 } } \} < y_l e^{ - \epsilon_1 }, $ for
$l=1,2$. Since $ x_l \in [ \frac{ \epsilon^{2}_1 }{ e^{ \epsilon_1 } }  -
\delta , \frac{ \epsilon^{2}_1 }{ e^{ \epsilon_1 } }  +
\delta ] $, it suffices to verify that
$
\epsilon^{1}_1 + \delta e^{ \epsilon_1 } < \exp \left( - \epsilon^{1}_0 - \frac{ \epsilon^{1}_1  }{ \epsilon^{1}_1 - \delta e^{ \epsilon_1 }   }    \right)
,
$
for an appropriate choice of $ \delta > 0 $ and $
\epsilon^{1}_0.$
By continuity, it suffices to prove this inequality for $ \delta = 0$ and $
\epsilon^1_0 = 0$, which becomes $ \epsilon^{1}_1 e < 1, $ and follows from the assumptions imposed on $\epsilon^{1}_1$. $\qed$

%\begin{figure}
% \includegraphics[scale=0.8]{plot1.eps}
%\caption{Multiple equilibria}\label{plot}
%\end{figure}

\subsection{Non-uniqueness with vanishing endowments}
In Theorem \ref{het_equil1} we assumed that $ P (
\epsilon^{i}_k > 0 ) = 1 $, for all $k=1,...,T$ and all
$i=1,...,N$. This assumption was crucial for proving that every
equilibrium SPD is of the form $(\ref{spd})$. It
turns out that once this assumption is relaxed, there
necessarily exist infinitely many equilibria, all of which of the same canonical
form.

\begin{theo}
Assume that $ P( \epsilon_k = 0 ) > 0$, for $k=0,...,T$ and
$P ( \cup_{k=0}^{T} \{ \epsilon^{i}_k > 0 \}) > 0,$ for  $i=1,...,N$. Then,
there exist infinitely many equilibria. Every equilibrium SPD
$( \widetilde{\xi}_k)_{ k = 0 , ... ,T } $ is of the form
\begin{equation}
\widetilde{\xi}_k (\widetilde{\lambda}_1 , ... , \widetilde{\lambda}_N  )  =  \xi_k( \widetilde{\lambda}_1 , ... , \widetilde{\lambda}_N )
\mathbf{1}_{ \{ \epsilon_k \neq  0 \} }  +  X_k \mathbf{1}_{ \{ \epsilon_k = 0 \} } ,
\label{infinity}
\end{equation}
for all $k=1,...,T$, where $\xi_k( \widetilde{\lambda}_1 , ... ,
\widetilde{\lambda}_N ) $ is given by (\ref{spd}) and $ X_k $ is
some non-negative $\mathcal{F}_k$-measurable random variable.
The constants $\widetilde{\lambda}_1 , ... , \widetilde{\lambda}_N $ are
determined by the budget constraints
\[
\sum_{k=0}^{T} \left(  E \left[ \widetilde{\xi}_k (\lambda_1 , ... , \lambda_N  ) ) I_{i} \left( \lambda_i e^{\rho_i k } \widetilde{\xi}_k ( \lambda_1 , ... , \lambda_N )  \right)   \right] - E \left[ \widetilde{\xi}_k( \lambda_1 , ... , \lambda_N )  \epsilon^{i}_k  \right] \right) = 0
\]
for $i=1,...,N$.
\label{non-uniqueness}
\end{theo}
${}$
\newline
\textbf{Proof of Theorem \ref{non-uniqueness}.} The proof is
identical to the proof of Theorem \ref{het_equil1} apart from a slight modification as follows.
Consider equation (\ref{modif}) and note that in the current
context this equation admits the form $ \mathbf{1}_{ \{
\widetilde{\xi}_k > \beta_{i_N (k)} (k)\}} = 1 $ on the set $ \{
\epsilon_k = 0\}$, which implies that $\widetilde{\xi}_k$ is of
the form (\ref{infinity}). $\qed $
\newline
\newline
We illustrate the above result through the following elementary
example.
\newline
\newline
\textbf{Example: Infinitely Many Equilibria.} Let ($\Omega$, $ \mathcal{F}_{1}$, $P$) be a probability space where
 $ \Omega = \{ \omega_1, \omega_2 \}$, $ P ( \{ \omega_1 \} ),
P ( \{ \omega_2 \} ) > 0$, $ \mathcal{F}_0 = \{ \Omega , \emptyset
\} $ and $ \mathcal{F}_1 = 2^{ \Omega }.$ Consider a one period
homogeneous economy with an individual represented by the utility
function $u(x) = -e^{-x} $ and $\rho=0.$ The endowments of the
agent are denoted by $\epsilon_0$ and $ \epsilon_1$. For the sake
of transparency, we analyze the following two simple cases
directly by using the definition of equilibrium rather than by
using Theorem \ref{non-uniqueness}.
\newline
(i) Let $ \epsilon_0 = 0 $ and $ \epsilon_1  $ be an arbitrary
$\mathcal{F}_1-$measurable positive random variable. Theorem
\ref{general_probability_space} implies that the optimal
consumption policies $c_0$ and $ c_1$ are given by $ c_0 = -  \log
\left( \mathbf{1}_{ \left\{ \lambda > 1 \right\} } + \lambda
\mathbf{1}_{ \left\{ \lambda \leq 1  \right\} }  \right) $ and $
c_1 =  - \log \left( \mathbf{1}_{ \left\{ \lambda \xi_1 > 1
\right\} } + \lambda \xi_1 \mathbf{1}_{ \left\{ \lambda \xi_1 \leq
1 \right\} } \right) $. The market clearing condition $c_0 = 0$
and $ c_1 = \epsilon_1 $ implies that $\xi_1 = \frac{ e^{ -
\epsilon_1 } }{ \lambda  } $ is an equilibrium SPD, for all $ \lambda \geq 1.$ Note that the budget
constraints of the type (\ref{constraint}) are evidently
satisfied. We stress that for the
corresponding unconstrained problem
\[
\sup_{ (c_0  , c_1 )  } -
e^{ -c_0 }  - E [ e^{-c_1} ],
\]
where $c_0 \in \mathbb{R}$ and $c_1 \in L^{2} \left( \mathcal{F}_1 \right)$ are such that
$c_0 + E [\xi_1 c_1 ] = \epsilon_0 + E [ \xi_1 \epsilon_1] ,$
there exists a unique equilibrium corresponding to $ \lambda = 1,$
that is, $ \xi_1 = e^{ - \epsilon_1 }.$
\newline
(ii) Let $ \epsilon_0 > 0 $ be arbitrary, $ \epsilon_1( \omega_1) > 0
$ and $ \epsilon_1( \omega_2) = 0. $ Then, by Theorem
\ref{general_probability_space} we obtain that $c_0 = \log( 1 /
\lambda) = \epsilon_0 $ and $c_1 = \max \{ \log ( \frac{1}{
\lambda \xi_1 }  ) , 0 \} = \epsilon_1 $. It follows that $
\lambda = e^{ - \epsilon_0 } $, and that there are infinitely many
equilibrium state price densities of the form $ \xi_1 ( y ) = e^{
\epsilon_0 - \epsilon_1 } \mathbf{1}_{ \{  \epsilon_1 \neq 0  \} }
+ y \mathbf{1}_{ \{  \epsilon_1 = 0  \} } $, one for every $ y > e^{
\epsilon_0 }.$ $\qed$

\section{Long-run yields of zero coupon bonds}
In this section we work with a complete market and an \textit{infinite} time horizon ($T=\infty$). We emphasize that all our
results in the context of equilibrium hold in this setting, due to a specific choice of the aggregate endowment process
(see (\ref{end_1})).
Recall that $\xi_k = \xi_k( \lambda_1 , ... , \lambda_N ),$ $k \in \mathbb{N}$ is the equilibrium SPD (see (\ref{spd})).
The equilibrium price at time 0 of a \textit{zero coupon bond} maturing at period $t \in \mathbb{N}$ is defined by
\begin{equation}
B^t = E \left[ \xi_t \right].
\label{bonds_def1}
\end{equation}
Based on Cramer's large deviation theorem, we study the asymptotic behavior (as $ t \to \infty$)
of the \textit{yield} at time 0, defined by
\begin{equation}
Y(0,t) := - \frac{ \log B^t}{t} .
\label{bonds_def2}
\end{equation}
The weights $ \lambda_1 , ... , \lambda_N $ are omitted here since they have no impact on the latter limit.
Throughout this section we assume that the total endowment process in the economy is
a \textit{random walk with drift}, i.e.,
\begin{equation}
\epsilon_k = \sum_{j=1}^{k} X_j,
\label{end_1}
\end{equation}
for all $k \in \mathbb{N},$ where $X_0=0$ and $X_1, X_2, ...$ are non-negative i.i.d random variables with a
finite mean $E[X_1]>0$.
%This specification of endowments is quite natural in this setting, due to the exponential preferences
%of the investors. We consider two categories of economies: agents with heterogeneous risk-aversion and agents with heterogeneous discounting rates.

\subsection{Heterogeneous risk-aversion}
Consider an economy where agents differ only with respect to the
risk-aversion, that is, $\gamma_1 < ... <  \gamma_N $, $\epsilon^{i}_k = \epsilon_k / N,$
$i=1,...,N$, $k \in \mathbb{N}$; $\rho_1 = ... = \rho_N =\rho $.
Recall (\ref{beta}) and note that in the present setting, equations (\ref{numbers}) can be rewritten as
\begin{equation}
\sum_{k=0}^{\infty} E \left[ \xi_k  \left( \log \left( \beta_i(k) \right) - \log \xi_k \right)^{+} \right]
 = \frac{\gamma_i}{N} \sum_{k=0}^{\infty} E \left[ \xi_k  \epsilon_k \right] .
\label{eq_A}
\end{equation}
Now, let $i , j \in \{ 1, ...  , N\} $ be arbitrary. Notice that the homogeneity of the impatience rate among agents implies
that either $\beta_i(k) \leq \beta_j(k),$ or $\beta_i(k) \geq \beta_j(k) $, for all $k \in \mathbb{N}$. Therefore, by (\ref{eq_A})
we conclude that $ \beta_1(k) < ... < \beta_N (k) $ for all $k \in \mathbb{N}$ and thus $ i_l(k) = l,$ for all $l=1,...,N.$
Hence, we get (see (\ref{eta}))
\[
 \eta_j = \log \left( \left( \frac{ \lambda_{j} }{ \gamma_{j}}  \right)^{ 1 / \gamma_{j+1} + ... + 1 / \gamma_{N}  }
\left(  \frac{ \gamma_{j+1} }{ \lambda_{j+1} } \right)^{ 1 / \gamma_{j+1} }  ... \left(
\frac{ \gamma_{N} }{ \lambda_{N} } \right)^{1 / \gamma_N}
  \right) ,
\]
for all $j=1,...,N$, and (see (\ref{spd}))
\[
 \xi_k = \sum_{j=1}^{N} \prod_{l=j}^{N}
\left( \frac{ \gamma_l }{ \lambda_l e^{\rho k} } \right)^{ \left( \sum_{ m= j }^{ N } \frac{\gamma_l}{ \gamma_m} \right)^{-1} }
\exp \left( - \frac{ \epsilon_k }{ \sum_{l=j}^{N} 1 / \gamma_l}  \right) 1_{ \{ \eta_j \leq  \epsilon_k < \eta_{j-1} \} },
\]
for all $k \in \mathbb{N}.$

\begin{theo}
In economies with heterogeneous risk-aversions, we have
\[
\lim_{t \to \infty} - \frac{ \log B^t }{ t }
=
\rho
- \log
E \left[
\exp \left( - \frac{ X_1 }{ \sum_{l=1}^{N} 1 / \gamma_l }  \right)
\right].
\]
\label{th_1}
\end{theo}
\textbf{Proof of Theorem \ref{th_1}.}
First, we evidently have
\[
- \frac{ \log B^t}{t}
\leq
-
\frac{ \log E \left[
\prod_{l=1}^{N}
\left( \frac{ \gamma_l }{ \lambda_l e^{\rho t} } \right)^{ \left( \sum_{ m= 1 }^{ N } \frac{\gamma_l}{ \gamma_m} \right)^{-1} }
\exp \left( - \frac{ \epsilon_t }{ \sum_{l=1}^{N} 1 / \gamma_l}  \right) 1_{ \{ \eta_1 \leq  \epsilon_t < \eta_{0} \} }
\right] }{t}.
\]
Next, note that the law of large numbers implies that
$ \lim_{t \to \infty } P \left( a \leq \epsilon_t \leq b \right) = 0 $, for all $0 \leq a \leq b,$
and $ \lim_{t \to \infty } P \left( c \leq \epsilon_t \right) = 1 $, for any $0 \leq c.$
Therefore, since $\eta_0=\infty$, we get
\[
-
\frac{ \log E \left[ N
\prod_{l=1}^{N}
\left( \frac{ \gamma_l }{ \lambda_l e^{\rho t} } \right)^{ \left( \sum_{ m= 1 }^{ N } \frac{\gamma_l}{ \gamma_m} \right)^{-1} }
\exp \left( - \frac{ \epsilon_t }{ \sum_{l=1}^{N} 1 / \gamma_l}  \right) 1_{ \{ \eta_1 \leq  \epsilon_t < \eta_{0} \} }
\right] }{t}
\leq
 - \frac{ \log B^t}{t} ,
\]
for sufficiently large $t$. It is left to prove that
\[
 \lim_{t \to \infty}
-
\frac{ \log E \left[
\exp \left( - \frac{ \epsilon_t }{ \sum_{l=1}^{N} 1 / \gamma_l}  \right) 1_{ \{ \eta_1 \leq  \epsilon_t < \eta_{0} \} }
\right] }{t} =
- \log
E \left[
\exp \left( - \frac{ X_1 }{ \sum_{l=1}^{N} 1 / \gamma_l }  \right)
\right].
\]
First, observe that

\begin{align*}
\lim_{t \to \infty}
-
\frac{ \log E \left[
\exp \left( - \frac{ \epsilon_t }{ \sum_{l=1}^{N} 1 / \gamma_l}  \right) 1_{ \{ \eta_1 \leq  \epsilon_t < \eta_{0} \} }
\right] }{t}
\geq
\lim_{t \to \infty}
-
\frac{ \log E \left[
\exp \left( - \frac{ \epsilon_t }{ \sum_{l=1}^{N} 1 / \gamma_l}  \right) \right] }{t}
\\
=
 - \log
E \left[
\exp \left( - \frac{ X_1 }{ \sum_{l=1}^{N} 1 / \gamma_l }  \right)
\right].
\end{align*}
On the other hand, we have
\[
-
\frac{ \log E \left[
\exp \left( - \frac{ \epsilon_t }{ \sum_{l=1}^{N} 1 / \gamma_l}  \right) 1_{ \{ \eta_1 \leq  \epsilon_t < \eta_{0} \} }
\right] }{t}
\]
\[
\leq
-
\frac{ \log E \left[
\exp \left( - \frac{ \epsilon_t }{ \sum_{l=1}^{N} 1 / \gamma_l}  \right) 1_{ \{ \frac{\eta_1}{t} \leq  X_1 \} }
... 1_{ \{ \frac{\eta_1 }{t} \leq  X_t \} }
\right] }{t}
\]
\[
=
-
\log E \left[
\exp \left( - \frac{ X_1 }{ \sum_{l=1}^{N} 1 / \gamma_l}  \right) 1_{ \{ \frac{\eta_1}{t} \leq  X_1 \} }
\right].
\]
Lastly, dominated convergence yields
\[
 \lim_{t \to \infty} -
\log E \left[
\exp \left( - \frac{ X_1 }{ \sum_{l=1}^{N} 1 / \gamma_l}  \right) 1_{ \{ \frac{\eta_1}{t} \leq  X_1 \} }
\right] =
-
\log E \left[
\exp \left( - \frac{ X_1 }{ \sum_{l=1}^{N} 1 / \gamma_l}  \right)
\right],
\]
completing the proof. $\qed$

\subsection{Heterogeneous impatience rates}
Consider an economy which is composed of agents who differ only with respect to the
impatience rates, that is, $\gamma_1= ... =  \gamma_N = \gamma$; $\epsilon^{i}_k = \epsilon_k / N,$
$i=1,...,N$, $k=1,2,...$; $\rho_N < ... < \rho_1$. Let $\lambda_1,...,\lambda_N$ be the weights corresponding to
the equilibrium SPD $\xi_k = \xi_k (\lambda_1 , ... , \lambda_N),$ $k \in \mathbb{N}$.
Note that there exists $t' \in \mathbb{N}$ such that $ \lambda_1 e^{\rho_1 t} > .... > \lambda_N e^{\rho_N t} ,$
for all $t>t'$. Therefore, by recalling (\ref{beta}) we get $i_{l}(t) = l $ for all $t>t',$ and consequently
(see (\ref{eta}) and (\ref{spd})), we have
\begin{equation}
 \eta_j(t) = \frac{1}{\gamma} \sum_{l=j+1}^{N} \log ( \beta_l / \beta_j ) =
\frac{1}{\gamma}  \sum_{l=j+1}^{N} \log ( \lambda_j / \lambda_l )  + \frac{1}{\gamma}  \sum_{l=j+1}^{N} ( \rho_j - \rho_l ) t ,
\label{rho_jt}
\end{equation}
for all $j=1,...,N,$ and
\[
 \xi_{t} =
 \gamma \sum_{j=1}^{N} (\lambda_j ... \lambda_N )^{-(N-j+1)^{-1}} \exp \left( - \frac{\rho_j + ... + \rho_N }{ N - j + 1 } t
 - \frac{ \gamma }{ N - j + 1 } \epsilon_t \right)
\]
\[
\times \textbf{1}_{ \{ \eta_j(t) \leq \epsilon_t < \eta_{j-1}(t) \} },
\]
for all $t>t'.$ Consider the logarithmic moment generating function of $X_1$
\[
 \Lambda (x ) = \log E \left[ e^{ \lambda X_1 } \right] ,
\]
and denote by
\[
 \Lambda^{*} (y ) = \sup_{ x \in \mathbb{R} } \left( x y - \Lambda ( y)    \right) ,
\]
the corresponding Legendre transform of $\Lambda.$ We set
\[
a_j = \rho_j + \inf_{x \in
 \left[ \frac{1}{\gamma} \sum_{l=j+1}^{N} (\rho_j - \rho_l )
, \frac{1}{\gamma} \sum_{l=j}^{N} (\rho_{j-1} - \rho_l ) \right]} \Lambda^{*}(x),
\]
\[
b_j = \rho_{j-1} + \inf_{x \in
 \left( \frac{1}{\gamma} \sum_{l=j+1}^{N} (\rho_j - \rho_l )
, \frac{1}{\gamma} \sum_{l=j}^{N} (\rho_{j-1} - \rho_l ) \right)} \Lambda^{*}(x),
\]
for $j=2,...,N,$ and
\[
a_1 =
\rho_1 + \inf_{x \in
 \left[ \frac{1}{\gamma} \sum_{l=2}^{N} (\rho_1 - \rho_l )
,
\infty \right)
} \Lambda^{*}(x).
\]
We are ready to state the main result of this subsection.

\begin{theo}
In economies with heterogeneous impatience rates, we have
\begin{equation}
\limsup_{t \to \infty} - \frac{ \log B^t}{t} \leq
\min \left\{ b_2 , ... , b_N \right\}
,
\label{limsup1}
\end{equation}
and
\begin{equation}
\liminf_{t \to \infty} - \frac{ \log B^t}{t} \geq
\min \left\{ a_1 , ... , a_N \right\}.
\label{liminf1}
\end{equation}
\label{theo_2}
\end{theo}

\textbf{Proof of Theorem \ref{theo_2}.}
Observe that the following inequality is satisfied
\[
\xi_{t}  \leq
 \gamma \sum_{j=1}^{N} (\lambda_j ... \lambda_N )^{- ( N - j + 1 )^{-1} }
\exp \left( -\frac{1}{ N - j + 1 } \left( \left( \rho_j + ... + \rho_N \right) t + \gamma \eta_{j}(t)  \right)   \right)
\]
\[
 \times
\textbf{1}_{ \{ \eta_{j}(t) \leq \epsilon_t < \eta_{j-1}(t) \} } \leq
\gamma \sum_{j=1}^{N}
( \lambda_j )^{-1} \exp \left( - \rho_j t \right)
 \textbf{1}_{ \{ \eta_{j}(t) \leq \epsilon_t < \eta_{j-1}(t) \} } ,
\]
for all $t>t'$. Denote $ a_j(t):= ( \lambda_j )^{-1} \exp \left( - \rho_j t \right)
 \textbf{1}_{ \{ \eta_{j}(t) \leq \epsilon_t < \eta_{j-1}(t) \} } $. We have
\begin{align*}
 \liminf_{t \to \infty}  - \frac{ \log B^t }{t}  &\geq
 \liminf_{t \to \infty}  - \frac{ \log E [ \sum_{j=1}^{N} a_j(t) ] }{ t }
\\
&\geq
 \liminf_{t \to \infty} \log \left(  \frac{1}{ \left( N \max \{ E \left[ a_1(t) \right] , ... , E \left[ a_N(t) \right] \} \right)^{1/t} } \right)
\\
&=
\liminf_{t \to \infty} \log \left(  \frac{1}{ \left( \max \{ E \left[ a_1(t) \right] , ... , E \left[ a_N(t) \right] \} \right)^{1/t} } \right)
\\
&=
 \liminf_{t \to \infty} \min \left\{ \log \left(  \frac{1}{ \left( E \left[ a_1(t) \right] \right)^{1/t} } \right)  , ... ,
\log \left(  \frac{1}{ \left( E \left[ a_N(t) \right] \right)^{1/t} } \right)
\right\}
\\
&=
\min \left\{ \liminf_{t \to \infty} \log \left(  \frac{1}{ \left( E \left[ a_1(t) \right] \right)^{1/t} } \right)  , ... ,
\liminf_{t \to \infty} \log \left(  \frac{1}{ \left( E \left[ a_N(t) \right] \right)^{1/t} } \right)
\right\}.
\end{align*}
Fix an arbitrary $\varepsilon>0.$ Next, recall (\ref{rho_jt}) and observe that the following
inequality holds true for each $j \in \{ 1,...,N \} $.
\begin{equation}
 \liminf_{t \to \infty} - \frac{ \log E \left[ a_j(t) \right] }{t } = \rho_j + \liminf_{t \to \infty} -
\frac{ \log P \left( \eta_{j}(t) \leq \epsilon_t < \eta_{j-1}(t) \right) }{ t }
\label{est_1}
\end{equation}
%\[
% \rho_j + \liminf_{t \to \infty} -
%\frac{ \log P \left(  \frac{1}{\gamma} \sum_{l=j+1}^{N} (\rho_j - \rho_l )
%+ \frac{1}{t} \frac{1}{\gamma} \sum_{l=j+1}^{N} \log \frac{\lambda_j}{\lambda_l}
% \leq \epsilon_t <
%\frac{1}{\gamma} \sum_{l=j}^{N} (\rho_{j-1} - \rho_l )
%+ \frac{1}{t} \frac{1}{\gamma} \sum_{l=j}^{N} \log \frac{\lambda_{j-1}}{\lambda_l}
% \right) }{ t } \geq
%\]
\[
\geq \rho_j + \liminf_{t \to \infty} -
\frac{ \log P \left(  \frac{1}{\gamma} \sum_{l=j+1}^{N} (\rho_j - \rho_l )
- \varepsilon
 \leq \epsilon_t \leq
\frac{1}{\gamma} \sum_{l=j}^{N} (\rho_{j-1} - \rho_l )
+ \varepsilon
 \right) }{ t }
\]
\[
\geq
 \rho_j + \inf_{x \in
 \left[ \frac{1}{\gamma} \sum_{l=j+1}^{N} (\rho_j - \rho_l )
- \varepsilon ,
\frac{1}{\gamma} \sum_{l=j}^{N} (\rho_{j-1} - \rho_l )
+ \varepsilon \right]
} \Lambda^{*}(x),
\]
where in the last inequality we have employed Cramer's large deviation theorem (see e.g. Theorem 2.2.3 in Dembo and Zeitouni (1998)).
Since $\varepsilon > 0$ was arbitrary, we conclude that
\[
\liminf_{t \to \infty} - \frac{ \log E [ a_j(t) ]  }{t } \geq
 \rho_j + \inf_{x \in
 \left[ \frac{1}{\gamma} \sum_{l=j+1}^{N} (\rho_j - \rho_l )
,
\frac{1}{\gamma} \sum_{l=j}^{N} (\rho_{j-1} - \rho_l )
 \right]
} \Lambda^{*}(x).
\]
The preceding inequalities combined with (\ref{est_1}) prove the validity of (\ref{liminf1}).
On the other hand, we have
\[
  \xi_{t} ( \lambda_1 , ... , \lambda_N ) \geq
\]
\[
 \gamma \sum_{j=2}^{N} (\lambda_j ... \lambda_N )^{- ( N - j + 1 )^{-1} }
\exp \left( -\frac{1}{ N - j + 1 } \left( \left( \rho_j + ... + \rho_N \right) t + \gamma \eta_{j-1}(t)  \right)   \right)
\]
\[
\times
 \textbf{1}_{ \{ \eta_{j}(t) \leq \epsilon_t < \eta_{j-1}(t) \} } \geq
 \gamma \sum_{j=2}^{N}
( \lambda_{j-1} )^{-1} \exp \left( - \rho_{j-1} t \right)
 \textbf{1}_{ \{ \eta_{j}(t) \leq \epsilon_t < \eta_{j-1}(t) \} } .
\]
Finally, inequality (\ref{limsup1}) follows analogously. $\qed$

\section{Precautionary savings}
This section deals with \textit{precautionary savings},
namely, savings resulted by future uncertainty. When markets are complete, one
would not anticipate this phenomenon to occur, since in essence, all risks can be hedged.
Thus, we concentrate on incomplete markets.
Here, as commonly referred to in the literature (see e.g. Carroll and Kimball (2008)), \textit{savings} should be understood literally
as less consumption. We consider a one-period incomplete market of type $\mathcal{C}$, as in Subsection 3.2.1, and stick to the same notation.
The only distinction is the particular specification of the random endowments.
Let $X \in L^{2} \left( \mathcal{F}_1 \right)$ be an arbitrary non-negative random variable. Denote by
\begin{equation}
\epsilon_1 = \epsilon_1( \varepsilon ) := \frac{ e^{ \varepsilon X } }{ E \left[ e^{ \varepsilon X }  \big| \mathcal{H}_1 \right]},
\label{eps1}
\end{equation}
the endowment of the agent at time $T=1$, for some $\varepsilon \in [0,1].$ Note that
$E \left[ \epsilon_1 (\varepsilon) \big| \mathcal{H}_1 \right]$ $ =1,$
and $\epsilon_1=1$ in case that $X \in L^{2} \left( \mathcal{H}_1 \right) $ or in case that the market is complete (i.e., if $\mathcal{F}_1 =
\mathcal{H}_1$). As shown in the next statement, the variance of $\epsilon_1 ( \varepsilon) $ is
an increasing function of $\varepsilon.$
\begin{Lemma}
The conditional variance on $\mathcal{H}_1$ of the random variable $\epsilon_1( \varepsilon )$ is an increasing function of $\varepsilon$, namely, the function
\[
 Var \left[ \epsilon_1(\varepsilon) \big| \mathcal{H}_1 \right] =
E \left[ \left( \epsilon_1(\varepsilon) - E \left[ \epsilon_1(\varepsilon) \big| \mathcal{H}_1 \right]   \right)^2   \big| \mathcal{H}_1 \right],
\]
satisfies the inequality
\[
Var \left[ \epsilon_1(\varepsilon_1) \big| \mathcal{H}_1 \right]
 \leq
 Var \left[ \epsilon_1(\varepsilon_2) \big| \mathcal{H}_1 \right] ,
\]
$P-a.s.,$ for all $ \varepsilon_1 \leq \varepsilon_2.$
\label{variance}
\end{Lemma}

\textbf{Proof of Lemma \ref{variance}.}
First note that
\[
Var \left[ \epsilon_1(\varepsilon) \big| \mathcal{H}_1 \right] = \frac{ E \left[ e^{2 \varepsilon X }  \big| \mathcal{H}_1  \right] }{
\left( E \left[ e^{ \varepsilon X }  \big| \mathcal{H}_1  \right]  \right)^2 } - 1.
\]
To complete the proof, it suffices to show that the corresponding differential is non-negative.
One checks that it holds if and only if
\[
 E \left[ X e^{ 2 \varepsilon X } | \mathcal{H}_1 \right]  E \left[ e^{ \varepsilon X }  | \mathcal{H}_1 \right] -
E \left[ e^{ 2 \varepsilon X } | \mathcal{H}_1 \right] E \left[ X e^{ \varepsilon X }  | \mathcal{H}_1 \right] \geq 0.
\]
To see this, consider the measure $Q$ defined by the Radon-Nykodym derivative
\[
 \frac{dQ}{dP} = \frac{ e^{2 \varepsilon X } }{ E \left[ e^{2 \varepsilon X } | \mathcal{H}_1 \right] },
\]
and note that the above inequality is equivalent to
\[
 E^{Q} \left[ X \big| \mathcal{H}_1 \right] E^{Q} \left[ e^{ - \varepsilon X } \big| \mathcal{H}_1 \right]  \geq
E^{Q} \left[ X e^{ - \varepsilon X } \big| \mathcal{H}_1 \right].
\]
The latter inequality follows from the FKG (see Fortuin et al. (1971)) inequality, since it can be rephrased as
$$
E^{Q} \left[ f(X) \big| \mathcal{H}_1 \right] E^{Q} \left[ g(X)  \big| \mathcal{H}_1 \right]  \geq E^{Q} \left[ f(X) g(X) \big| \mathcal{H}_1 \right],
$$
where $f(x) = x$ is increasing and $g(x) = e^{ - \varepsilon x }$ is decreasing. $\qed$
\newline
\newline
According to Lemma \ref{variance}, the parameter $\varepsilon \in [0,1]$ quantifies the degree of income uncertainty.
More precisely, an increase in $\varepsilon$ yields an increase in the variance of the income, but preserves the mean. Therefore,
we learn that the un-insurability of income is an increasing function of the parameter $\varepsilon$.
Recall the explicit formulas (see (\ref{c0_incomplete}) and (\ref{c1_incomplete})) for the optimal consumption stream
in the current setting. One can show (similarly to Corollary \ref{corollary}) that by choosing a sufficiently large $\epsilon_0$ (the initial endowment),
the corresponding optimal consumption stream coincides with the unconstrained one, and it is given by
%\[
%\widehat{c}_0 = \frac{1 }{ \gamma_i} \left(
%\log \left( \frac{1 }{ \lambda }   \right) + \rho_i
%\right)^{+},
%\]
%and
%\[
%\widehat{c}_1 =
%\frac{1}{\gamma_i} \log \left( \frac{ e^{\gamma_i \epsilon^{i}_1}  E \left[ e^{ - \gamma_i \epsilon^{i}_1 } \big| \mathcal{H}_1 \right] }{ \lambda M_1 } \right)
%1_{ \{ essinf [ e^{\gamma_i \epsilon^{i}_1} | \mathcal{H}_1 ] E
%[ e^{ - \gamma_i \epsilon^{i}_1} | \mathcal{H}_1 ]
%> \lambda M_1 \} }
%+
%\]
%\[
% \frac{1}{\gamma_i}
%\log \left( \frac{ e^{ \gamma_{i} \epsilon^{i}_1 } }{ essinf [ e^{  \gamma_{i} \epsilon^{i}_1 }  | \mathcal{H}_1 ] } \right)
%1_{ \{
%essinf [ e^{\gamma_{i} \epsilon^{i}_1} | \mathcal{H}_1 ] E
%[ e^{ - \gamma_{i} \epsilon^{i}_1} | \mathcal{H}_1 ]
% \leq \lambda M_1 \} }.
%\]
%and recall that the constant $\lambda$ is determined uniquely by the equation $ \widehat{c}_0 +
%E \left[ M_1 \widehat{c}_1 \right] = \epsilon_0 + E \left[ M_1 \epsilon_1( \varepsilon) \right]. $
%Observe that the finiteness of the probability space and the fact that $\varepsilon \in (0,1]$ yields
%the existence of a constant $K$ such that for every choice of $\epsilon_0 > K,$ we will have
%\[
%e^{ - \gamma_i \epsilon^{i}_1} | \mathcal{H}_1 ]
%> \lambda M_1,
%\]
%$P-$a.s., and thus in particular $ \lambda < 1.$ In particular, the optimal consumption streams coincide with the
%unconstrained ones and are given by
\begin{equation}
\widehat{c}_0 ( \varepsilon ) = \frac{1 }{ \gamma} \left(
\log \left( \frac{1 }{ \lambda ( \varepsilon  ) }   \right) + \rho
\right),
\label{prec0}
\end{equation}
and
\begin{equation}
\widehat{c}_1 ( \varepsilon )
= \frac{1}{\gamma} \log \left( \frac{ e^{\gamma \epsilon_1 ( \varepsilon ) }
E \left[ e^{ - \gamma \epsilon_1 ( \varepsilon ) } \big| \mathcal{H}_1 \right] }{ M_1 } \right)
+ \widehat{c}_0 ( \varepsilon ) - \frac{1}{\gamma} \rho,
%\frac{1}{\gamma} \log \left( \frac{ e^{\gamma \epsilon_1 ( \varepsilon ) }  E \left[ e^{ - \gamma \epsilon_1 ( \varepsilon ) }
%\big| \mathcal{H}_1 \right] }{ \lambda ( \varepsilon ) M_1 } \right)
\label{prec1}
\end{equation}
%\[
%= \frac{1}{\gamma} \log \left( \frac{ e^{\gamma \epsilon_1 ( \varepsilon ) }
%E \left[ e^{ - \gamma \epsilon_1 ( \varepsilon ) } \big| \mathcal{H}_1 \right] }{ M_1 } \right)
%+ \widehat{c}_0 ( \varepsilon ) - \frac{1}{\gamma} \rho
%\]
where $\lambda ( \varepsilon ) $ (or equivalently, $ \widehat{c}_0 ( \varepsilon ) $) solves uniquely the equation
\begin{equation}
\widehat{c}_0 ( \varepsilon ) + E \left[ M_1 \widehat{c}_1 ( \varepsilon ) \right] = \epsilon_0 + E \left[ M_1 \epsilon_1( \varepsilon) \right].
\label{prec3}
\end{equation}
We are ready to state the main result of the section.
\begin{theo}
The precautionary savings motive holds true. Namely,
$ \widehat{c}_0 ( \varepsilon )$ is a decreasing function of $\varepsilon$. Hence, in particular,
investors consume less in the present, as the variance of future-income increases.
\label{prec_sav}
\end{theo}\textbf{Proof of Theorem \ref{prec_sav}.}
First, it is possible to rewrite constraint (\ref{prec3})
as
\[
\widehat{c}_0(\varepsilon) \left( 1 + E \left[ M_1 \right] \right) +
\frac{1}{\gamma} E \left[ M_1 \log E \left[ e^{ - \gamma \epsilon_1  ( \varepsilon) }  \big| \mathcal{H}_1 \right] \right]
= K ,\]
where $K:= \frac{\rho}{\gamma} E \left[ M_1 \right] + \epsilon_0 + \frac{1}{\gamma} E \left[ M_1 \log M_1 \right]$
is a constant not depending on $\varepsilon.$ Obviously, $\widehat{c}_0(\varepsilon)$ is a differentiable function of
$\varepsilon.$ We get
\[
\frac{ \partial \widehat{c}_0 }{ \partial \varepsilon} =
\frac{1}{1 + E \left[ M_1 \right]} E \left[ M_1 \frac{
E \left[ e^{ - \gamma \epsilon_1 ( \varepsilon ) } \frac{ \partial \epsilon_1}{\partial \varepsilon}  \big| \mathcal{H}_1 \right]
}{ E \left[ e^{ - \gamma \epsilon_1 ( \varepsilon ) }  \big| \mathcal{H}_1 \right] }   \right],
\]
where the differential of $\epsilon_1(\varepsilon)$ is given by (see (\ref{eps1}))
\[
\frac{ \partial \epsilon_1}{\partial \varepsilon} =
\frac{X e^{\varepsilon X } }{ E \left[ e^{ \varepsilon X } \big| \mathcal{H}_1 \right] }
-
\frac{  e^{\varepsilon X } E \left[ X e^{\varepsilon X } \big| \mathcal{H}_1 \right] }{
\left( E \left[ e^{ \varepsilon X } \big| \mathcal{H}_1 \right] \right)^2 }
.
\]
Therefore, to complete the proof, it suffices to check that
$
E \left[ e^{ - \gamma \epsilon_1 ( \varepsilon ) } \frac{ \partial \epsilon_1}{\partial \varepsilon}  \big| \mathcal{H}_1 \right] \leq 0.
$
This boils down to proving
\[
  E \left[ X e^{\varepsilon X } e^{ - \gamma \epsilon_1 (\varepsilon) } \big| \mathcal{H}_1 \right]
E \left[ e^{ \varepsilon X } \big| \mathcal{H}_1 \right]
\leq
 E \left[ e^{ - \gamma \epsilon_1 (\varepsilon) } e^{\varepsilon X } \big| \mathcal{H}_1 \right]
E \left[ X e^{\varepsilon X } \big| \mathcal{H}_1 \right].
\]
Set a new measure $Q$ given by the Radon-Nykodym derivative
$
\frac{dQ}{dP} :=
\frac{ e^{ \varepsilon X } }{ E \left[ e^{\varepsilon X } \big| \mathcal{H}_1 \right] } ,
$
and note that by dividing the preceding inequality by $\left( E \left[ e^{ \varepsilon X } \big| \mathcal{H}_1 \right] \right)^2$,
we arrive at
\[
 E^{Q} \left[ X  e^{ - \gamma \epsilon_1 (\varepsilon) } \big| \mathcal{H}_1 \right]
\leq
 E^{Q} \left[ e^{ - \gamma \epsilon_1 (\varepsilon) } \big| \mathcal{H}_1 \right]  E^Q \left[ X  \big| \mathcal{H}_1 \right].
\]
By (\ref{eps1}), we have
$
 X = \left( \log \epsilon_1 (\varepsilon) + \log E \left[ e^{\varepsilon X } \big| \mathcal{H}_1 \right] \right) \frac{1}{\varepsilon}.
$
Therefore, the required inequality admits the form
\[
 E^{Q} \left[ e^{ - \gamma \epsilon_1 (\varepsilon) } \log \epsilon_1 (\varepsilon) \big| \mathcal{H}_1  \right]
\leq
E^{Q} \left[ e^{ - \gamma \epsilon_1 (\varepsilon) }  \big| \mathcal{H}_1  \right]
E^{Q} \left[ \log \epsilon_1 (\varepsilon) \big| \mathcal{H}_1  \right],
\]
which follows from the FKG inequality, as in Lemma \ref{variance}. $\qed$

\end{document}